\documentclass[11pt]{article}
\pdfoutput=1
\usepackage[utf8]{inputenc}
\usepackage{amsmath,amsfonts,amssymb}
\usepackage{amsthm}
\usepackage{latexsym}
\usepackage[noadjust]{cite}
\usepackage{graphicx}
\usepackage{xcolor}
\usepackage{dirtytalk}
\usepackage{mathtools}
\usepackage[margin=1in]{geometry}
\usepackage{thm-restate}
\usepackage{enumitem}
\usepackage[linesnumbered,ruled]{algorithm2e}
\usepackage[colorlinks=true, allcolors=blue]{hyperref}
\usepackage[nameinlink,capitalise]{cleveref}
\hypersetup{
    citecolor={violet}
}

\newtheorem{theorem}{Theorem}[section]

\newtheorem{lemma}[theorem]{Lemma}

\newtheorem{claim}[theorem]{Claim}
\theoremstyle{definition}
\newtheorem{definition}[theorem]{Definition}
\newtheorem{remark}[theorem]{Remark}
\newtheorem{fact}[theorem]{Fact}

\newcommand{\E}{\mathbb{E}}

\newcommand{\Z}{\mathbb{Z}}

\newcommand{\rank}{\text{rank}}

\newcommand{\eps}{\epsilon}

\newcommand{\M}{\mathcal{M}}
\newcommand{\I}{\mathcal{I}}
\newcommand{\Ind}{\mathrm{Ind}}
\newcommand{\Bin}{\mathrm{Bin}}
\newcommand{\Hyp}{\mathrm{Hyp}}
\newcommand{\Var}{\mathbf{Var}}

\title{On the Parallel Complexity of Finding a Matroid Basis}
\date{\today}

\author{Sanjeev Khanna\thanks{School of Engineering and Applied Sciences, University of Pennsylvania, Philadelphia, PA. Supported in part by NSF award CCF-2402284 and AFOSR award FA9550-25-1-0107. Email: {\tt sanjeev@cis.upenn.edu}.} \and Aaron Putterman\thanks{School of Engineering and Applied Sciences, Harvard University, Cambridge, Massachusetts, USA. Supported in part by the Simons Investigator Awards of Madhu Sudan and Salil Vadhan, NSF Award CCF 2152413 and AFOSR award FA9550-25-1-0112. Email: \texttt{aputterman@g.harvard.edu}.} \and Junkai Song\thanks{School of Engineering and Applied Sciences, University of Pennsylvania, Philadelphia, PA. Email: \texttt{junkais@cis.upenn.edu}.}}

\begin{document}

\maketitle

\begin{abstract}

A fundamental question in parallel computation, posed by Karp, Upfal, and Wigderson (FOCS 1985, JCSS 1988), asks: \emph{given only independence-oracle access to a matroid on $n$ elements, how many rounds are required to find a basis using only polynomially many queries?} This question generalizes, among others, the complexity of finding bases of linear spaces, partition matroids, and spanning forests in graphs.
In their work, they established an upper bound of $O(\sqrt{n})$ rounds and a lower bound of $\widetilde{\Omega}(n^{1/3})$ rounds for this problem, and these bounds have remained unimproved since then.

In this work, we make the first progress in narrowing this gap by designing a parallel algorithm that finds a basis of an arbitrary matroid in $\tilde{O}(n^{7/15})$ rounds (using polynomially many independence queries per round) with high probability, surpassing the long-standing $O(\sqrt{n})$ barrier. Our approach introduces a novel matroid decomposition technique and other structural insights that not only yield this general result but also lead to a much improved new algorithm for the class of \emph{partition matroids} (which underlies the $\widetilde\Omega(n^{1/3})$ lower bound of Karp, Upfal, and Wigderson). Specifically, we develop an $\tilde{O}(n^{1/3})$-round algorithm, thereby settling the round complexity of finding a basis in partition matroids. 

As a further application, we also improve the parallel complexity of the classic \emph{matroid intersection} problem. By plugging our basis-finding algorithm into a known algorithmic framework for matroid intersection, we obtain an $\tilde{O}(n^{37/45})$ round algorithm for matroid intersection, improving upon the prior $O(n^{5/6})$ bound.

Collectively, these results represent the first progress on the parallel complexity of finding matroid bases in 40 years, and we believe that techniques developed here may prove useful for other problems on matroids. 
\end{abstract}

\vspace{3cm}

\pagenumbering{gobble}

\pagebreak

\tableofcontents
\pagebreak

\pagenumbering{arabic}

\section{Introduction}

A central pursuit in algorithm design is to understand which problems admit efficient parallel solutions. A common and well-studied measure is the \emph{round complexity}: how many rounds of adaptive queries (or steps) are required to solve a problem, where each round can perform polynomially many operations/queries in parallel. This model has been extensively studied in both theoretical and applied settings, and has yielded parallel algorithms for a range of fundamental problems---including maximal independent sets~\cite{Lub86}, matchings in graphs~\cite{Lov79, KUW86, FGT16, ST17}, submodular function minimization~\cite{BS20, CCK21, CGJS22} and matroid intersection~\cite{GGR22, GT17, Bli22,BT25}.

In this work, we focus on parallel computation in matroids. Matroids are a powerful abstraction capturing the structure of independence in many combinatorial settings, including forests in graphs, linearly independent vectors, and feasibility in optimization problems. We revisit a foundational open problem posed by Karp, Upfal, and Wigderson~\cite{kuw85, KUW88}:

\begin{quote}
\emph{Given oracle access to a matroid $\M$ on $n$ elements, how many adaptive rounds are required to find a basis of $\M$?}
\end{quote}

We assume the algorithm has access only to an \emph{independence oracle}, which takes a subset $S \subseteq E$ and returns whether $S \in \mathcal{I}$, where $\mathcal{I}$ is the collection of independent sets. This is the most general model of matroid access and captures the full generality of matroid theory. In this setting, no further structure is assumed; even linear or graphic representations of the matroid are unavailable.

The importance of this question stems in part from the broad applicability of matroids. For example, when $\M$ is a graphic matroid, the problem reduces to finding a spanning forest of a graph using only queries to a cycle oracle. For linear matroids, it becomes finding a basis of a subspace without direct access to coordinates. Given the super-exponential number of matroids~\cite{BPV15}, understanding the round complexity in this general oracle model is a natural and foundational challenge.

A \emph{matroid} $\M = (E, \mathcal{I})$ consists of a ground set $E$ and a family of subsets $\mathcal{I} \subseteq 2^E$ satisfying three axioms: (1) $\emptyset \in \mathcal{I}$; (2) hereditary property ($S \subseteq T \in \mathcal{I} \Rightarrow S \in \mathcal{I}$); and (3) exchange property (if $S, T \in \mathcal{I}$ with $|S| < |T|$, there exists $e \in T \setminus S$ such that $S \cup \{e\} \in \mathcal{I}$). A \emph{basis} is a maximal independent set; all bases of a matroid have the same size, called the \emph{rank} of $\M$.

In their foundational work, Karp, Upfal, and Wigderson~\cite{kuw85} gave a parallel algorithm that finds a basis in $O(\sqrt{n})$ adaptive rounds using polynomially many independence queries per round. They also gave an $\widetilde{\Omega}(n^{1/3})$\footnote{Throughout the paper, we use $\widetilde{O}(\cdot)$ and $\widetilde\Omega(\cdot)$ to hide factors of $\mathrm{polylog}(\cdot)$.} round lower bound for a family of \emph{partition matroids}, showing that any algorithm using polynomially many queries per round must use at least $\widetilde\Omega(n^{1/3})$ rounds. These bounds have remained the best known for nearly forty years.

\subsection{Our Contributions}

We make the first progress on narrowing the gap between upper and lower bounds established in the work of Karp, Upfal, and Wigderson~\cite{kuw85}. Our main result is a faster randomized algorithm that breaks the $\Theta(\sqrt{n})$ round barrier:

\begin{theorem}[Main Result]\label{thm:introMain}
There is a randomized algorithm that, with high probability, finds a basis of any $n$-element matroid in $\tilde{O}(n^{7/15})$ adaptive rounds, using only polynomially many independence queries.
\end{theorem}

\begin{remark}
As an immediate consequence, given a \emph{weighted} matroid (where every element has an associated weight), there is also an $\tilde{O}(n^{7/15})$ round algorithm for finding a maximum (or minimum) weight basis, as shown in \cite{BT25}. 
\end{remark}

Our algorithm relies on several new techniques and structural results, including, a \emph{matroid decomposition framework} that partitions the ground set into a small number of ``components'' that are amenable to localized processing; a new characterization of \emph{rank deficiency} in terms of random sampling; and a parallel routine for \emph{recovering redundant elements} that improves over previous deletion-based strategies.

Our techniques (in fact, a simplified version of them) also lead to significantly improved round complexity for the important special case of {\em partition matroids}.

\paragraph{Partition matroids.} Partition matroids form a natural and widely studied subclass of matroids and were used in the lower bound construction of~\cite{kuw85}. We show that the known $\widetilde\Omega(n^{1/3})$ lower bound is essentially tight for this class by providing a matching algorithm:

\begin{theorem}\label{thm:introPartition}
There is a {\em deterministic} algorithm that finds a basis of any $n$-element partition matroid in $\tilde{O}(n^{1/3})$ adaptive rounds, using only polynomially many independence queries.
\end{theorem}

\paragraph{Matroid intersection.} The matroid intersection problem is a common generalization of several fundamental combinatorial optimization problems, including bipartite matching, arborescence in directed graphs, and the colorful spanning tree problem. Recent work by~\cite{BT25} showed that matroid intersection can be solved via calls to a matroid basis oracle. Plugging in our improved basis-finding algorithm immediately gives a better round complexity (improving upon \cite{BT25}'s $O(n^{5/6})$ rounds):

\begin{theorem}\label{thm:matroidIntersectionIntro}
There is a randomized algorithm that, with high probability, finds a maximum (weight) common independent set of two $n$-element matroids in $\tilde{O}(n^{37/45})$ adaptive rounds, using only polynomially many independence queries.
\end{theorem}

We now give an overview of the main technical ideas underlying our results.

\subsection{Technical Overview}

\subsubsection{Prior Work}

First, we recap the algorithm and analysis (for the upper bound) of \cite{kuw85}. We start by reviewing two natural operations on matroids:

\begin{enumerate}
    \item \textbf{Contraction:} Contraction in a matroid $\M = (E, \I)$ refers to finding a set of independent elements $S$, and committing to include these in our basis. The contracted matroid (denoted $\M / S$) has elements $E \setminus S$, and a set $T \subseteq E \setminus S$ is independent \emph{if and only if} $T \cup S$ is independent in $\M$. For this reason, when we contract on a set $S$, we can still simulate independence oracle queries to $\M / S$ by simply appending the set $S$ to the query (and then querying $\M$). Importantly, when we contract on a set $S$, the remaining number of elements in $\M/ S$ decreases by $|S|$, and so in this sense, the problem of finding a basis in $\M$ is reduced to that of finding a basis in a smaller instance.
    \item \textbf{Deleting Redundant Elements:} The second method of making ``progress'' towards recovering a basis in a matroid is by deleting redundant elements. Specifically, we say that a set $S$ of elements in a matroid $\M$ is \emph{redundant}, if $\mathrm{rank}(\M\setminus S) = \mathrm{rank}(\M)$. Clearly, if we can find such a redundant set, then finding a basis in $\M$ reduces to finding a basis in $\M \setminus S$. 
\end{enumerate}

Now, we describe the $\mathrm{poly}(n)$-query $O(\sqrt{n})$-round algorithm for finding bases in arbitrary matroids as first presented in \cite{kuw85}. With the previous notions already established, the algorithm itself is quite simple: given a matroid $\M$ on $n$ elements $e_1, \dots e_n$, the algorithm first splits the elements up into $\sqrt{n}$ groups: $S_1 = \{e_1, \dots e_{\sqrt{n}} \}$, $S_2 = \{e_{\sqrt{n} + 1}, \dots e_{2\sqrt{n}} \}$, $\dots$, $S_{\sqrt{n}} = \{e_{n - \sqrt{n}+1}, \dots e_{n} \}$. Now, the queries that the algorithm makes are simply all prefixes of $S_i$, for every $i \in [\sqrt{n}]$ (i.e., for $S_1$, the queries would be $\{e_1\}, \{e_1, e_2 \},  \{e_1, e_2, e_3\} \dots \{e_1, e_2, \dots e_{\sqrt{n}}\}$). There are only two cases for us to consider:
\begin{enumerate}
    \item If, for any set $S_i$, we see that $\mathrm{Ind}(S_i) = 1$ (i.e., the entire set is independent), then the algorithm simply contracts on $S_i$. In particular, this means that we will have recovered an independent set of size $\sqrt{n}$, and so when we contract on the set $S_i$, we see that $\mathrm{Rank}(\M / S_i) = \mathrm{Rank}(\M) - \sqrt{n}$. Thus, in future rounds, we simply operate on the new matroid defined by $\M / S_i$.
    \item If there are no sets $S_i$ for which $\mathrm{Ind}(S_i) = 1$, this means that every set $S_i$ was dependent. In particular, because we queried every prefix of $S_i$, we can also identify the first query which became dependent. I.e., there were two queries where $\{e_{(i-1) * \sqrt{n} + 1}, \dots e_{(i-1) * \sqrt{n} + j}\}$ was independent, but $\{e_{(i-1) * \sqrt{n} + 1}, \dots e_{(i-1) * \sqrt{n} + j+1}\}$ was dependent. A simple fact in matroid theory states that when this happens, $e_{(i-1) * \sqrt{n} + j+1}$ is a \emph{redundant} element for any matroid which contains elements $\{e_{(i-1) * \sqrt{n} + 1}, \dots e_{(i-1) * \sqrt{n} + j}\}$, and can therefore be deleted. Thus, summing across all $\sqrt{n}$ groups, we can delete $\sqrt{n}$ redundant elements.
\end{enumerate}

Thus, we see that in either case we are making progress towards recovering a basis of the matroid. Either we recover $\sqrt{n}$ independent elements towards a basis, or we delete $\sqrt{n}$ redundant elements. In both cases, we reduce the problem to finding a basis on a matroid with $ n - \sqrt{n}$ elements, and so a simple recursive calculation reveals that this terminates in $O(\sqrt{n})$ rounds. 

\subsubsection{Warming Up with Partition Matroids}\label{sec:introPartitionMatroids}

In the work of \cite{kuw85}, their lower bound of $\widetilde\Omega(n^{1/3})$ rounds for finding a basis relied on the class of \emph{partition matroids}. In this setting, the matroid $\M = (E, \I)$ has its elements $E$ partitioned across $k$ parts, denoted $P_1, \dots P_k$. Each set $P_i$ is also given a \emph{budget} $b_i \in \Z^{\geq 0}$. Given a set $S \subseteq E$, we define the rank as 
\[
\mathrm{rank}(S) = \sum_{i = 1}^k \min(b_i, |S \cap P_i|).
\]
Essentially, as we add elements to a given part, its rank continues to increase until it reaches its budget, at which point the rank remains fixed. Note for a given part $P_i$, once we find a set of $b_i+1$ elements that belongs to $P_i$, we can in fact \emph{completely} recover $P_i$ in a single additional round. If we denote such a set by $A_i$, this is because we can remove a single element $x$ from $A_i$, and then query the independence oracle with $(A_i \setminus \{x\}) \cup \{y\}$ for every other element $y \in \M$. The only elements which will make such a query dependent are those that are also in the part $P_i$.

Given this definition, it is actually quite straightforward to define the lower bound instances that \cite{kuw85} relied on. For an $n$ element ground set, they create $n^{1/3}$ parts, denoted $P_1, \dots P_{n^{1/3}}$, where each part contains $n^{2/3}$ elements, and the budget of the $i$-th part is exactly $i \cdot n^{1/3}$. This defines the \emph{structure} of the matroid, but the actual assignment of labels to the elements (i.e., which elements are in which part) is decided uniformly at random. In the first round now, because the labels are decided uniformly at random, any query of elements is effectively a random sample of the $n$ elements of the matroid (and, for notation, let us say the query is denoted by $B$, and the random sampling is performed at some rate $\beta = |B| / n$). 

The key point is that the response to the query $B$ is \emph{with extremely high probability} completely governed by the elements in $B \cap P_1$. Indeed, if the sampling rate $\beta \geq \frac{1.5}{n^{1/3}}$, then with extremely high probability, the query $B$ is dependent, as $>(1 - \eps) \cdot \beta \cdot n^{2/3} > n^{1/3}$ elements will be sampled from $P_1$ which exceeds the budget $b_1$. Likewise, if $\beta < \frac{1.5}{n^{1/3}}$, then the budgets among the sets $P_2, \dots P_{n^{1/3}}$ are \emph{not} exceeded, and so the independence or dependence of the query $B$ is dictated entirely by $B \cap P_1$. Intuitively then, this means that in the first round of queries, the only information that is being leaked relates to the set of elements $P_1$. The argument can then be repeated, where in the second round, the only leaked information is from $P_2$, and so on. This argument is formalized in \cite{kuw85}.

However, the above construction is very suggestive: could we hope for a better lower bound using partition matroids? In fact, a natural starting point for this question would be to consider a partition matroid with $\sqrt{n}$ parts $P_1, \dots P_{\sqrt{n}}$, where each part has $\sqrt{n}$ elements, and the budget $b_i$ of part $P_i$ is $i$. If an induction argument as used in the previous setting held true, then we could even hope for an $\Omega(\sqrt{n})$ lower bound, matching the algorithm provided by \cite{kuw85}. 

It turns out however, that this is not possible. The problem that arises is the following: suppose we have eliminated the first $i-1$ parts, so now the matroid consists of $P_i, \dots P_{\sqrt{n}}$. Next, consider a query $B$ which consists of sampling each element independently with probability approximately $\frac{i}{\sqrt{n}}$. In expectation, the number of elements that are sampled from each part is exactly $\frac{i}{\sqrt{n}} \cdot \sqrt{n} = i$. However, the problem that arises is that the number of surviving elements follows a binomial distribution and therefore \emph{anti-concentrates} with non-negligible probability. Indeed, there is a non-negligible probability that $P_i$ receives fewer than $i$ elements, while $P_{i+1}$ \emph{receives more} than $i+1$ elements, and thus it is actually $P_{i+1}$ which dictates the response to query $B$. In fact, because the standard deviation scales as $\sqrt{i}$, it is even the case that if we make the sampling rate slightly less than $\frac{i}{\sqrt{n}}$, \emph{any} of the parts $P_i, \dots P_{i + \Omega(\sqrt{i})}$ has a non-negligible chance of dictating the response to the query $B$. This allows for an algorithm to gain too much information about the underlying parts $P_i, \dots P_{i + \Omega(\sqrt{i})}$, and can in fact reveal all of their exact identities in a single round. Repeating this argument leads to an algorithm that solves this instance in $O(n^{1/4})$ rounds! 

Thus, while this thought experiment does not yield new lower bounds, it does provide an algorithmic insight, namely, instances of partition matroids where each adaptive round can only recover a single part, must have the property that the budgets are ``well-separated'' (by a standard deviation at least). This ensures that in such an  instance with $k$ parts, the budgets must grow to $\Omega(k^2)$. This means that after $n^{1/3}$ rounds of recovery we are dealing with a partition matroid with $\Omega(n^{2/3})$-size budget and hence $O(n^{1/3})$ remaining parts. It follows that the remaining parts can be discovered in $O(n^{1/3})$ additional rounds. This is essentially the basis for our $\tilde{O}(n^{1/3})$ round algorithm to find basis of any partition matroid. In fact, because the class of partition matroids is only exponentially large, we can even derandomize this algorithm. We do this by repeating our decomposition procedure $\mathrm{poly}(n)$ times in each round, thereby boosting the success probability to be $1 - 2^{-\mathrm{poly}(n)}$, at which point we can take a union bound over \emph{every possible partition matroid}. This implies that there is single, fixed polynomial size set of queries which simultaneously works for recovering a basis \emph{across all} partition matroids. This set of queries constitutes the deterministic algorithm.

However, all of the analysis above is heavily tailored to the setting of partition matroids. The next question is if these insights can be extended to {\em general} matroids, by developing analogues of the notions of parts and budgets. It turns out that the answer is a ``yes'' for many of the above notions if we develop a sampling-based view of these concepts. In the remainder of the technical overview, we introduce these new quantities and explain how they can be stitched together to improve on \cite{kuw85}'s long-standing upper bound. 

\subsubsection{Extending Notions from Partition Matroids to General Matroids}

To extend the algorithmic approach used for partition matroids to arbitrary matroids, we must address a fundamental question:
\begin{quote}
\emph{What is the appropriate analogue of a ``part'' in a general matroid, and how can we discover such structure via random sampling?}
\end{quote}

Our main tool is a new decomposition framework that simulates the process of ``peeling off'' components from the matroid, in analogy to how we recovered parts in the partition matroid case. Instead of relying on explicit part definitions, we use random permutations to expose dependence structure. The core idea is that, under random sampling, elements or sets that frequently cause early dependence can be interpreted as playing the role of ``tight'' components.

\paragraph{Circuits from random prefixes.}

Let $\M = (E, \mathcal{I})$ be a matroid of rank $r$. Consider a random permutation $\pi$ of the ground set $E$. 
Define a prefix process as before: let $S_j := \{\pi(1), \dots, \pi(j)\}$, and find the smallest index $j^*$ such that $S_{j^*}$ is dependent. Because $j^*$ is the smallest such index, $S_{j^*-1}$ is independent, and the dependent set $S_{j^*}$ contains a unique minimal dependent subset, i.e., a \emph{circuit} $C_\pi \subseteq S_{j^*}$. 
In particular, we can actually \emph{exactly} recover the elements in this circuit that has formed: If we query with $S_{j^*}  - \{x\}$ for $x \in C_\pi$, then this query will in fact be independent. Likewise, if we query with $S_{j^*}  - \{x\}$ for $x \notin C_\pi$, the resulting query is dependent, as the circuit $C_\pi$ is still contained in $S^{\pi}_{j}$. 

We will view the circuit $C_\pi$ as a random variable: it is the unique first circuit encountered over the randomness of $\pi$. We now define the following notions:
\begin{itemize}
    \item For a subset $S \subseteq E$, let $q_\M(S) := \Pr_\pi[C_\pi \subseteq S]$ be the probability that the first circuit lies entirely in $S$.
    \item For an element $x \in E$, let $p_\M(x) := \Pr_\pi[x \in C_\pi]$ be the probability that $x$ appears in the first circuit.
\end{itemize}

We can interpret $p_\M(x)$ as the ``circuit participation mass'' of element $x$ under random prefix sampling.

\paragraph{Greedily-optimal sets.}

We now define the building blocks of our decomposition, namely, \emph{greedily-optimal sets}. Informally speaking, a subset $S \subseteq E$ is a greedily-optimal set if it maximizes $q_\M(S)$ (the probability that the first circuit lies inside $S$), subject to a stability condition: removing any element from $S$ significantly reduces $q_\M(S)$. 

\begin{definition}
    We say that a set $S^* \subseteq E$ is \emph{greedily-optimal} if 
    \[
    q_{\M}(S^*) \geq 1-2^{-20} + \frac{\sum_{i = 1}^{|S^*|} \frac{1}{i}}{2^{20}\log n},
    \]
    and there is no element $x \in S^*$ such that 
    \[
    q_{\M}(S^*\setminus\{x\}) \geq 1-2^{-20} + \frac{\sum_{i = 1}^{|S^*|-1} \frac{1}{i}}{2^{20}\log n}.
    \]
    Note that here $2^{20}$ is just a sufficiently large constant to ensure our probabilistic arguments go through.
\end{definition}

There are two key useful properties of greedily-optimal sets. First, they can be found efficiently. This is because we can start with the complete set $S^* = E$, and as long as there exists an element $x$ which satisfies the second condition, we update $S^* \leftarrow S^* - \{x \}$. All that we need to know for this are the probabilities $q_\M (S)$ for every $S$, and we can estimate these to very high accuracy by empirically evaluating them on a large sample of random permutations (and this takes only a single round). Second, once we find a greedily optimal set $S^*$, this means $q_{\M}(S^*) \geq 1-2^{-20} + \frac{\sum_{i = 1}^{|S^*|} \frac{1}{i}}{2^{20}\log n}$, but for any element we delete, $q_{\M}(S^* - \{ x \}) < 1-2^{20} + \frac{\sum_{i = 1}^{|S^*|-1} \frac{1}{i}}{2^{20}\log n}$. Importantly, this means for every element $x \in S^*$, it intuitively participates in a large fraction of the circuits that appear. Formally, we let $p_{\M|_{S^*}}(x) = \Pr_{\pi}[x \in C_{\pi}]$, where (now $C_{\pi}$ refers to the circuits that appear when sampling a permutation restricted to $S^*$, \emph{not} the original matroid $\M$). We show that, in a very strong sense, this second piece of intuition is true:

\begin{claim}
    Let $\M=(E,\I)$ be a matroid on $n$ elements, and let $S^*$ be a greedily-optimal set. Then, for every $x \in S^*$,
    \[
    p_{\M|_{S^*}}(x) \geq \frac{1}{2\cdot 2^{20}|S^*| \log n}.
    \]
\end{claim}

These greedily-optimal sets will play the role of ``components'' in our general decomposition, mimicking the parts $P_i$ in partition matroids. While a greedily-optimal set $S$ does not come with an explicit budget, it can be characterized by an analogous notion using a parameter $\alpha(S)$ that measures how early dependences appear in $S$ under random sampling.

\paragraph{The $\alpha(S)$ parameter.}

For a set $S \subseteq E$, we define the parameter $\alpha(S)$ as the median prefix length (within $S$) required to trigger dependence:
\[
\alpha(S) := \mathrm{median}_{\pi} \left\{ j : \text{$\{\pi(1), \dots, \pi(j)\} \subseteq S$ is dependent, $\{\pi(1), \dots, \pi(j-1)\} \subseteq S$ is independent} \right\},
\]
where $\pi$ is a random permutation of $S$.

Informally, $\alpha(S)$ captures how ``tight'' the set $S$ is: smaller $\alpha(S)$ implies dependence tends to occur early when sampling from $S$. This plays the role of the ``budget'' in partition matroids: if $|S| = \ell$ and $\alpha(S) \approx b + 1$, then the behavior of $S$ resembles a part of size $\ell$ and budget $b$.

We use this parameter to stratify components of the matroid and guide their ordering in the decomposition.

\subsubsection{Building a Matroid Decomposition}

Our decomposition does the following: given a matroid $\M$, it finds a greedily optimal set $S_1$ of $\M$ in a single round. It peels this set off, and then repeats the decomposition starting with the matroid $\M \setminus S_1$. This procedure continues repeating until the matroid is exhausted (has no elements remaining). Note that a greedily-optimal set cannot be empty, so we are also guaranteed that this procedure terminates in a finite number of steps. We denote the resulting greedily-optimal sets that are peeled off by $S_1, \dots S_k$. 

The informal claim below helps bound the number of sets that we peel off and allows us to relate these sets to
one another.

\begin{claim}\label{clm:quadraticIntro}[Informal]
    Let $\M$ be a matroid, and let $S_i, S_j$ be any two sets that are peeled off in the course of the above decomposition, with $j > i$. Then,
    \[
    \alpha(S_j) = \frac{\alpha(S_i) |S_j|}{|S_i|} + \Omega \left (\sqrt{\frac{\alpha(S_i) |S_j|}{|S_i|}} \right ).
    \]
\end{claim}

To start, we give a brief sketch which describes why the above claim is true. Let $S_i, S_j$ be given, and let $\M_{i-1}$ denote the state of the matroid $\M$ after $i-1$ iterations. Then, $S_i$ is a greedily-optimal set with respect to $\M_{i-1}$. In particular, because $S_i$ is greedily-optimal, $q_{\M}(S_i) \geq 1- 1 / 2^{20}$: this means that when we sample a random permutation $\pi$ and add elements from $\M_{i-1}$ until there is a circuit, there is a $\geq 1- 1 / 2^{20}$ probability that the resulting circuit is completely contained in $S_i$. That is to say, it is exceedingly rare for there \emph{ever} to be a circuit which includes elements from outside $S_i$.

Now, instead of sampling a random permutation $\pi$ and adding elements in this order, we consider a slightly different procedure: we let $p = \frac{\alpha(S_i)}{|S_i|}$, and we consider what happens when we sample the matroid $\M_{i-1}$ at rate $p$. 
\begin{enumerate}
    \item Within the set $S_i$, we expect exactly $\alpha(S_i)$ elements to survive. In fact, with probability $\Omega(1)$, there are even $\leq \alpha(S_i) - 1$ elements selected from $S_i$. Because $\alpha(S_i)$ is defined as the median number of samples needed before a circuit appears in $S_i$, if $\leq \alpha(S_i) - 1$ elements are sampled from $S_i$, then with probability $\geq 1/2$, no circuit forms completely inside $S_i$. 
    \item At the same time, we consider what happens when we sample the set $S_j$. We expect $\frac{\alpha(S_i) |S_j|}{|S_i|}$ elements to be sampled from $S_i$. Indeed, because the number of elements sampled follows a binomial distribution, we even know that the standard deviation of the number of sampled elements is $\sqrt{\frac{\alpha(S_i) |S_j|}{|S_i|}}$. Thus, with constant probability, there are 
    \[
    \geq \frac{\alpha(S_i) |S_j|}{|S_i|} + \Omega \left ( \sqrt{\frac{\alpha(S_i) |S_j|}{|S_i|}}\right )
    \]
    elements that survive sampling in $S_j$.
\end{enumerate}

Now, we are ready to prove the above claim: when we sample $\M_{i-1}$ at rate $p$, there is a constant probability of both (1) no circuit in $S_i$, and (2) at least $\frac{\alpha(S_i) |S_j|}{|S_i|} + \Omega \left ( \sqrt{\frac{\alpha(S_i) |S_j|}{|S_i|}}\right )$ elements sampled from $S_j$. If we assume for the sake of contradiction that $\alpha(S_j) = \frac{\alpha(S_i) |S_j|}{|S_i|} + o \left (\sqrt{\frac{\alpha(S_i) |S_j|}{|S_i|}} \right )$, then point (2) would imply that there is also a constant probability of there being a circuit completely contained in $S_j$. However, this would yield a contradiction, as it implies that there is a constant probability of recovering a circuit in $S_j$ \emph{before} we recover a circuit in $S_i$. But, by our definition of $S_i$ being greedily optimal, we know that when we add random elements, a $\geq 1 - 1 / 2^{20}$ fraction of circuits appear only in $S_i$. The above argument would otherwise suggest that there is a larger than $1 / 2^{20}$ probability of circuits appearing \emph{outside} $S_i$.

Having established \cref{clm:quadraticIntro}, we can already observe some interesting behavior about our decomposition. For instance, if we have two sets $|S_i| = |S_j|$, then $\alpha(S_j) = \alpha(S_i) + \Omega(\sqrt{\alpha(S_i)})$. If we generalize this to more than just two sets, and instead suppose we have $|S_1| = |S_2| = \dots = |S_{\ell}|$, then we get a chain of growth in the alpha values:
\[
\alpha(S_2) = \alpha(S_1) + \Omega\left(\sqrt{\alpha(S_1)}\right ), \quad \alpha(S_3) = \alpha(S_2) + \Omega\left(\sqrt{\alpha(S_2)}\right ),
\]
and so on. Ultimately, this means that if we have $\ell$ sets of the same size, the $\alpha$ value of the final set is $\Omega(\ell^2)$. 

In fact, we can generalize this argument to sets whose sizes are within constant factors of one another:
\begin{lemma}
Let $\mathcal{M}$ be a matroid, and let $S_1, \dots S_k$ be a sequence of sets that are peeled off. Let $\ell\in [\log n]$ be an integer, let $T = \{ i \in [k]: |S_i| \in [2^{\ell}, 2^{\ell+1} -1]\}$, let $\gamma = |T|$, and let $a_{\gamma}$ be the largest index in $T$. Then it must be the case that 
\[
    \alpha(S_{a_{\gamma}}) = \Omega(\gamma^2).
\]
\end{lemma}
Likewise, because $\alpha(S) \leq |S|$, this also means that there can only ever be $O(\sqrt{2^{\ell}})$ sets whose sizes are in the range $[2^{\ell}, 2^{\ell+1} -1]$. This immediately gives us a bound on the number of sets that can be returned by our decomposition: first, there must be $\leq n^{1/3}$ sets of size $\geq n^{2/3}$ (otherwise all the elements of the matroid are removed). Now, for any $\ell \in [\frac{2}{3} \log n]$, there can only be $O(2^{\ell/2})$ sets of size $[2^{\ell}, 2^{\ell+1}]$. Summing over the values of $\ell$, the number of sets returned by the decomposition is bounded by $O(n^{1/3})$, exactly as we saw in the partition matroid case.

With these basic facts established about the decomposition, we now show how the decomposition gives us the power to make progress towards finding a basis in many different ways. 

\subsubsection{Making Progress through Contraction}

To simplify the above discussion, we consider an abridged version of the above decomposition, where instead of continuing until the matroid $\M$ is empty, we stop the decomposition as soon as there are $< n/2$ elements remaining. As before, we still denote the sets that are recovered by $S_1, \dots S_k$. Our key observation now is the following:

\begin{claim}\label{clm:introSubroutine1}
    In one additional round, we can recover an independent set of size 
    \[
    \max_{i \in [k]} \Omega \left ( \frac{\alpha(S_i)}{|S_i|} \cdot n \right ).
    \]
\end{claim}

Thus, if this ratio $\frac{\alpha(S_i)}{|S_i|}$ every becomes too large (i.e., imagine it becomes $\Omega(1)$), then in just a single additional round, we can recover many independent elements, thereby making progress towards recovering an independent set. 

To see why the claim is true, consider any matroid $\M$ along with a greedily optimal set $S$ of $\M$. As $S$ is greedily optimal, it should be the case that $S$ contains the vast majority of circuits that appear when random sampling. So if we sample $\M$ at rate $\frac{\alpha(S)}{10|S|}$, then we know that there is a constant probability of \emph{no} circuits in $S$, and therefore there must also not be any circuits in $\M$ (with high probability), which in turn implies the sampled set is independent. In order to amplify this success probability, we can just repeat this sampling procedure for $\text{poly}(n)$ times (in parallel) and obtain an independent set of size $\Omega\left(\frac{\alpha(S)}{|S|}|\M|\right)$ with high probability. To get the above claim in its exact form, we can simply repeat this procedure for \emph{every} $S_i$ that is peeled off; i.e., with the matroid $\M$ and $S_1$, with $\M_1 = (\M \setminus S_1)$ and $S_2$, and so on. Because we stopped the decomposition before $\M$ reaches $< n/2$ elements, in each case, the remaining elements have cardinality $|\M|\geq \Omega(n)$. This yields the claim.

On an intuitive level, the benefit from being able to contract is that we can now assume that $\frac{\alpha(S_i)}{|S_i|} = o(1)$, for if not, there is a simple mechanism for making progress. Even better, since our only objective is to beat $n^{0.5}$ rounds, we can even assume the gap is non-trivially large, some $n^{\eps}$ for $\eps > 0$. In the next section, we will show how we can make progress towards recovering redundant elements when this ``$\alpha$-gap'' is large. 

\subsubsection{Making Progress Through Explicit Solving}

As discussed in the previous section, we will assume that $\frac{\alpha(S_i)}{S_i} = o(1)$ is significantly smaller than $1$. We are motivated by the following observation: if, in the course of doing our decomposition, there are many of the sets $S_1, \dots S_k $ that are of size $[\tau/2, \tau]$, then we can in fact \emph{explicitly} find bases of these sets by investing $O(\sqrt{\tau})$ extra rounds, using the algorithm of \cite{kuw85}. Unfortunately, as already remarked by \cite{kuw85}, there is no guarantee that \emph{combining} bases together is making meaningful progress towards finding a basis of the entire matroid. Instead, we have to argue that each one of these sets has many \emph{redundant} elements that we can delete \emph{in parallel} to make progress. 

If we were only guaranteed that $\frac{\alpha(S_i)}{S_i} = o(1)$, it is hard to argue that there are many redundant elements to be deleted (consider for instance matroid that has $n^{\eps}$ elements with rank $0$). It is here where we first use the second key property of a greedily-optimal set: i.e., that \emph{every element} in the set $S_i$ has a large probability (approximately $\widetilde{\Omega}(1 / |S_i|)$) of being in the first recovered circuit when we randomly sample elements from $S_i$. It turns out that these two conditions suffice for non-trivially bounding the number of redundant elements in each $S_i$:

\begin{theorem}\label{thm:introSubroutine2}
    Let $S$ be a set of elements in a matroid $\M$ such that $\mathrm{rank}(S) =r$ and
    \begin{enumerate}
        \item $\alpha(S) \leq \frac{|S|}{100\log |S|}$.
        \item For every element $x \in E$, $p_{\M|_{S}}(x) \geq \frac{1}{|S|^{10}}$.
    \end{enumerate}
    Then, $|S| - r = \Omega(|S| / \log|S|)$.
\end{theorem}

To see why this theorem is true, we need several additional tools and pieces of notation: we let $\M' = \M|_S$ be the matroid with only elements in $S$, and we let $(\M')^*$ denote the dual matroid of $\M'$. This is the matroid on the same set of elements, whose bases are the \emph{complements} of bases in $\M'$, and whose rank is therefore $|S| - \mathrm{rank}(S)$. Next, we require the notion of a \emph{quotient} of a matroid: for a set $R \subseteq \M'$, we say that the quotient of $R$ (denoted $Q(R)$) is the set of elements $\M' - \mathrm{span}(R)$, where $\mathrm{span}(R) = \{x \in \M': \mathrm{rank}(\{x \} \cup R) = \mathrm{rank}(R)\}$. A key fact in matroid theory \cite{Oxl06} is that if a set of elements $C$ forms a circuit in $\M'$, then the \emph{same} set of elements forms a quotient in $(\M')^{*}$. Finally, we require a useful theorem from the work of Quanrud \cite{Qua23}, which we invoke on the \emph{dual matroid} $(\M')^{*}$: for any parameter $d \in \Z^+$, there is a set $P$ of $\leq d \cdot \mathrm{rank}((\M')^{*})$ elements, such that in the matroid $(\M')^{*} \setminus P$, for any $\alpha \in \Z^+$, there are at most $|S|^{2 \alpha}$ quotients of size $\leq \alpha \cdot d$. For intuition, this result is a strengthening of the \emph{cut-counting bound} first derived in the work of Karger \cite{Kar93}. The quotients correspond to the cuts of a graph, and the elements we remove correspond to the \emph{small cuts} that we remove in order to make the minimum cut larger. With these results established, we are ready to give an outline of the proof of our theorem.

To start, recall that our goal is to show that $|S| - \mathrm{rank}(S)$ is large. We can immediately see that this value is exactly the rank of the dual matroid $(\M')^{*}$, and so it suffices to lower bound $\mathrm{rank}((\M')^{*})$. We then proceed by contradiction: if it happens to be the case that $\mathrm{rank}((\M')^{*}) = o(|S| / \log|S|)$, then by the result of \cite{Qua23}, for $d = 100 \log|S|$, there exists a set $P \subseteq (\M')^*$ of size $|P| \leq d \cdot \mathrm{rank}((\M')^{*}) = o(|S|)$, such that removing this set of elements leads to a ``quotient counting bound'' with parameter $d$ in the matroid $(\M')^{*}\setminus P$. Importantly, this counting bound implies that if we sample the elements of $(\M')^{*}\setminus P$ at rate $1/2$, then the probability that there is \emph{any quotient} in $(\M')^{*}\setminus P$ for which \emph{every element} is sampled is bounded by $2/|S|^{97}$. This is because there is an implicit tradeoff: smaller quotients are more likely to survive sampling, but there are fewer of them. As we increase the quotient size, there are more quotients, but their ``survival'' probability decreases proportionately. Note that by circuit-quotient duality this means that if we sample the matroid $\left ( (\M')^{*}\setminus P \right )^*$ at rate $1/2$ we expect no \emph{circuits} to survive sampling. 

To reach a contradiction, we have to understand this matroid $\left ( (\M')^{*}\setminus P \right )^*$. A well-known property of matroids (see, for instance \cite{Oxl06}), is that deleting elements is actually \emph{dual} to contracting on elements. Thus, the matroid $\left ( (\M')^{*}\setminus P \right )^* = \left ( (\M')^{*} \right )^* / P = \M' / P$ (i.e., our starting matroid $\M'$ contracted on $P$).  By the previous paragraph, this implies that when we sample $\M' / P$ at rate $1/2$, with extremely high probability, there are no circuits that survive the sampling. However, using assumption (2) of our theorem statement, for any element $x \notin P$, we can show that sampling $\M'$ at rate $1/2$ yields a circuit involving $x$ with probability $\geq 1/|S|^{10}$. After this contraction on $P$, it is in fact the case that these elements in $\M' \setminus P$ \emph{are only more likely} to form circuits when sampling at rate $1/2$ compared to in $\M'$, and thus this probability is still $\geq 1/|S|^{10}$. But, this yields our contradiction: our first bound says that with probability $\geq 1 - 2 / |S|^{97}$, no circuits appear at this sampling rate, while our second bound says that circuits do appear with probability $\geq 1 / |S|^{10}$. Therefore, it must be the case that $\mathrm{rank}((\M')^{*}) = \Omega(|S| / \log|S|)$, as we desire.

Note that with this theorem in hand, we are guaranteed that for each set $S_i$ of size $[\tau/2, \tau]$ that we explicitly find a basis for, we can essentially delete $\widetilde\Omega(\tau)$ redundant elements, thereby guaranteeing some form of progress. Unfortunately, this method has a key drawback, which is that we \emph{explicitly} solve for a basis, which requires investing even more rounds of adaptivity. In the next section, we present our final approach for deleting redundant elements which deletes fewer redundant elements, but also requires only a single round of adaptivity. 

\subsubsection{Efficiently Finding Redundant Elements}

As mentioned above, our final method for recovering redundant elements recovers fewer redundant elements, but does so in only a single round, in a sense forming the analog of the procedure for recovering elements in a single part as outlined in \cref{sec:introPartitionMatroids}. We encapsulate the behavior of this routine below:

\begin{lemma}\label{lem:introSubroutine34build}
Let $\M$ be a matroid, and $S$ be a greedily-optimal set. Then, there is a $1$-round algorithm which recovers $\widetilde\Omega \left(\min\left\{|S|,\frac{|S|^2}{\alpha(S)^2}\right\}\right)$ redundant elements.
\end{lemma}

We omit a complete proof, but include the intuition here. Recall that because $S$ is greedily-optimal, every element $x \in S$ satisfies $p_{\M|_S}(x) = \widetilde \Omega (1 / |S|)$ (i.e., they frequently appear in the first circuit that arises under random sampling). We create an even more fine-grained understanding: we define $p_{x, \ell}$ to be the probability that $x$ appears in the first circuit that arises under random sampling \emph{conditioned} on $x$ being the $\ell$-th element included. In particular, we can observe that if $\ell \gg \alpha(S)$, this probability is essentially $0$, as the first circuit will already have formed by the time we add $x$. We can also see that these probabilities are monotonely decreasing, as $x$ is only more likely to participate in the first circuit when it is added earlier. These two facts are enough to derive that $p_{x, 1} = \widetilde \Omega\left ( \frac{|S|}{\alpha(S)} \cdot p_{\M|_S}(x) \right ) = \widetilde \Omega\left ( \frac{|S|}{\alpha(S) \cdot |S|} \right )$.

Now, we consider the following simple process: we sample a random set $A_1$ of size approximately $100 \alpha(S) \log|S| \gg \alpha(S)$. Now, for each element $x \in S$, we query the independence oracle with sets $\{x \} \cup W$, for $W$ being every prefix of the set $A_1$. Because $A_1$ is a completely random set, we see that the probability $x$ participates in the first circuit as we add elements to $W$ is essentially $p_{x, 1}$. If $x$ does appear in a circuit with $W$, then $x$ is in fact a \emph{redundant} element conditioned on $A_1$. Thus, for our deletion procedure, we simply keep all the elements in $A_1$, and remove all the elements \emph{outside} $A_1$ that formed circuits with $A_1$.

However, instead of stopping here, we repeat this procedure multiple times: we sample sets $A_2, A_3, \dots$ and so on. Ultimately, we can only sample $\approx \frac{|S|}{\alpha(S)}$ many sets before \emph{every} element is in one of the $A_i$'s (this would be an issue, as we cannot remove the elements inside the $A_i$'s). But, this is enough to boost the deletion probability of each element $x$ to be approximately 
\[
\Omega \left ( \min \left (1,  p_{x, 1} \cdot \frac{|S|}{\alpha(S)}\right ) \right )= \Omega \left ( \min \left (1, \frac{|S|^2}{\alpha^2(S)|S|}\right ) \right ).
\]
Summing across all elements in $S$, this yields the deletion of $\Omega \left ( \min \left (|S|, \frac{|S|^2}{\alpha^2(S)}\right ) \right )$ elements, as we desired.

With this routine now established, we are ready to complete the proof of our overall algorithm. 

\subsubsection{Putting the Pieces Together}

As before, we consider running our decomposition procedure until there are $< n/2$ elements remaining. We let the recovered sets be denoted by $S_1, \dots S_k$. By a simple pigeonhole argument, we are also guaranteed that there is some choice of $\ell \in [\log n]$ for which there are $\geq \frac{k}{\log n}$ of the sets $S_1, \dots S_k$ are of size $[2^{\ell}, 2^{\ell + 1}-1]$. We let $\tau$ denote this value $2^{\ell}$, and let $\gamma$ denote this number of sets of size $[\tau, 2\tau]$. We let $T$ refer to the indices of the sets whose sizes are in this range, and we let $\beta=\tau\cdot (\max_{i\in[k]}\alpha(S_{i})/|S_{i}|)$. 

To summarize the above subsections, we have the following methods of making progress: 

\begin{enumerate}
    \item From \cref{clm:introSubroutine1}, we can invest a single additional round beyond those $k = \widetilde{O}(\gamma)$ invested for the decomposition, and find an independent set of size $\Omega(\frac{n\beta}{\tau})$.
    \item From \cref{thm:introSubroutine2}, we can invest $O(\sqrt{\tau})$ extra rounds and find $\gamma  \cdot \widetilde\Omega\left( \tau\right)$ redundant elements. 
    \item From \cref{lem:introSubroutine34build}, we can invest a single additional round and find $\gamma  \cdot \widetilde\Omega\left( \min \left ( \tau, \frac{\tau^2}{\beta^2} \right )\right)$ redundant elements.
    \item Lastly, also from \cref{lem:introSubroutine34build}, we can recover $\sum_{i = 1}^k \widetilde \Omega \left (|S_i|, \frac{\tau^2}{\beta^2} \right) = \widetilde \Omega\left( \min\left(n, \frac{\tau^2}{\beta^2}\right)\right)$ redundant elements in a single additional rounds.
\end{enumerate}

We summarize this below:

\begin{center}
\begin{tabular}{|c||c|c|}
\hline
& Progress & Round Complexity\\
\hline
\cref{clm:introSubroutine1} & $\widetilde\Omega\left(n\beta/\tau\right)$ & $\widetilde O(\gamma)$ \\ 
\hline
\cref{thm:introSubroutine2} & $\widetilde\Omega\left(\gamma\tau\right)$ & $\widetilde O(\gamma+\tau^{1/2})=\widetilde O(\tau^{1/2})$ \\ 
\hline
\cref{lem:introSubroutine34build} & $\widetilde\Omega\left(\gamma \cdot \min\left(\tau,\frac{\tau^2}{\beta^2}\right)\right)$ & $\widetilde O(\gamma)$ \\ 
\hline
\cref{lem:introSubroutine34build} & $\widetilde\Omega\left(\min\left(n,\frac{\tau^2}{\beta^2}\right)\right)$ & $\widetilde O(\gamma)$ \\ 
\hline
\end{tabular}
\end{center}

To obtain \cref{thm:introMain}, we simply do a case analysis based on $\beta, \tau$ and $\gamma$. We show that \emph{for any} setting of these parameters, there is a choice of one of the above sub-routines which guarantees a Progress to Round ratio of at least $\widetilde \Omega(n^{8/15})$. If we let $\kappa$ denote the number of rounds invested in the sub-routine, this means the round complexity is governed by the recurrence $T(n) = \kappa + T(n - \kappa \cdot \widetilde \Omega(n^{8/15}))$, and a simple calculation shows then that the algorithm terminates in $\widetilde O(n^{7/15})$ rounds. 

\begin{remark}
    In fact, we can observe that \cref{thm:introSubroutine2} relies on using the $O(\sqrt{n})$ round algorithm of \cite{kuw85} to find bases of an arbitrary matroid. Now that we have an algorithm with complexity $\widetilde O(n^{7/15})$ rounds, we can actually \emph{improve} the round complexity of this sub-routine, and thereby achieve a better complexity than $\widetilde O(n^{7/15})$ rounds in the global algorithm. 
\end{remark}

\subsection{Organization}

In \cref{sec:prelim} we present some preliminary facts that we will make use of throughout our work. For the interested reader, in \cref{sec:partitionFull} we present a complete analysis deriving \cref{thm:introPartition} in a stand-alone manner. Subsequent sections of the paper focus on the general case, and can be read independently of \cref{sec:partitionFull}. In \cref{sec:decomp}, we present a formal analysis of our decomposition algorithm. In \cref{sec:largeAlpha}, we show how to make progress by recovering large independent sets. In \cref{sec:redundant}, we present a formal analysis of our techniques discussed above for recovering redundant elements. In \cref{sec:tradeoff}, we show how to trade-off between all of our subroutines for making progress to achieve our improved round complexity, thereby proving \cref{thm:introMain}.

\section{Preliminaries}\label{sec:prelim}

\subsection{Notation}

For a set $S$ and an element $x$, we let $S+x=S\cup\{x\}$ and $S-x = S\setminus \{x\}$ for short. We use $\binom{S}{i}$ to denote the set of all subsets of $S$ of cardinality $i$. For a matroid $\M=(E,\I)$ and $S\subseteq E$, we denote $\M|_{S}$ as the matroid restricted to set $S$, and $\M/S$ as the matroid after contracting $S$.

Throughout this paper, we denote $E_0$ as the ground set of the original matroid, $E$ as the ground set of current matroid, and let $n_0=|E_0|,n=|E|$. We always assume $n_0$ is sufficiently large.

\subsection{Matroid Theory}\label{sec:matroidTheoryPrelims}

\begin{definition}[Matroids]
A \emph{matroid} $\M=(E,\I)$ is a pair where $E$ is a finite ground set and $\I \subseteq 2^{E}$ is a collection of independent sets with the following properties: (i) $\emptyset \in \I$ (non-triviality), (ii) for every $S\in \I$ and $S'\subset S$, $S'\in \I$ (downward-closedness), and (iii) for every $S,S'\in \I$ and $|S'|<|S|$, there exists some $x\in S\setminus S'$ such that $S+x\in \I$ (exchange property).
\end{definition}

\begin{definition}[Independent Sets, Circuits, Bases]
For a matroid $\M=(E,\I)$, we say a set $S\subseteq E$ is \emph{independent} if $S\in \I$ and \emph{dependent} otherwise. We call a set $B$ a \emph{basis} if it is a maximal independent set, i.e. for any $x\notin B$, $B+x\notin \I$. We call a set $C$ a \emph{circuit} if it is a minimal dependent set, i.e. for any $x\in C$, $C-x\in \I$.
\end{definition}

We have the following fact:

\begin{fact}
\label{lem:matroid-circuit}
For a matroid $\M=(E,\I)$ and $S\subseteq E,x\in E\setminus S$, if $S\in \I$ and $S+x\not\in \I$, then there is a unique circuit $C$ in $S+x$ where $x\in C$. Moreover, for every $y\in C\setminus x$, $S-y+x\in \I$.
\end{fact}

\begin{definition}[Rank]
For a matroid $\M=(E,\I)$, we define the \emph{rank} of $\M$ as $\rank(\M) = \max_{S\in \I} |S|$. Further, for any $S\subseteq E$, we define $\rank_{\M}(S) = \max_{T\subseteq S,T\in \I} |T|$. The rank function of a matroid is submodular.
\end{definition}

\begin{definition}[Span, Flats]
In a matroid $\M=(E,\I)$, we define $\text{span}(S)$ as
\[
\text{span}(S) = \{x\in E \mid \text{rank}(S\cup \{x\})=\text{rank}(S)\}.
\]
We say that a set $S \subseteq E$ is a \emph{flat} if $S=\text{span}(S)$. Furthermore, when a set $S$ is a flat and satisfies $\mathrm{rank}(S) = \mathrm{rank}(\M) - 1$, we call $S$ a \emph{hyperplane}.
\end{definition}

\begin{definition}[Quotients]
Let $\M=(E,\I)$ be a matroid. Then, the set $Q = \{E\setminus S: S \text{ is a flat in }\M \}$ is called the \emph{set of quotients} in a matroid. 
\end{definition}

\begin{definition}[Dual Matroid]
Let $\M=(E,\I)$ be a matroid, the dual matroid $\M^* = (E,\I^*)$ is defined as
\[
\I^* = \{S\subseteq E\mid \exists B\subseteq(E\setminus S) \text{ s.t } B \text{ is a basis of }\M\}.
\]
In particular, $\rank(\M^*)=|E|-\rank(\M)$, and $(\M^*)^*=\M$.
\end{definition}

There are also the following useful facts about dual matroids:

\begin{fact}\cite{Oxl06}\label{fact:dualDeletion}
Let $\M=(E,\I)$ be a matroid, and $S\subseteq E$. Then,
\[
    (\M/S)^* = \M^* \setminus S.
\]
Furthermore, for all $T\subseteq E\setminus T$
\[
\rank_{\M/S} (T) = \rank_{\M}(T\cup S) - \rank_{\M}(S).
\]
\end{fact}

\begin{fact}\cite{Oxl06}\label{fact:circuitHyperplane}
Let $\M=(E,\I)$ be a matroid, and let $\M^*$ be its dual matroid. A set of elements $C \subseteq E$ is a circuit in $\M$ if and only if $E\setminus C$ is a hyperplane in $\M^*$.
\end{fact}

\subsection{Probability Theory}

In the analysis of our algorithm, we will need an anti-concentration bound for the hypergeometric distribution. We will approximate the hypergeometric distribution using a binomial distribution, for which the following anti-concentration bound is established in the literature:

\begin{lemma}[\cite{doe20}]
\label{lem:binomial}
Let $X\sim \Bin(n,p)$ with $p\leq 1/2, \Var[X]=np(1-p) \geq 1$, then
\[
\Pr\left[X\geq \E[X] + \frac{1}{5}\sqrt{\Var[X]}\right] \geq \frac{1}{108}, \quad \Pr\left[X\leq \E[X] - \frac{1}{5}\sqrt{\Var[X]}\right] \geq \frac{1}{108}
\]
\end{lemma}

We formally quantify the difference between the hypergeometric distribution and binomial distribution by bounding the total variation distance between them.

\begin{definition}[Total Variation Distance]
Let $P,Q$ be two probability distribution on $\mathbb{Z}$, the total variation distance between $P$ and $Q$ is defined as

\[
\delta(P,Q) = \sup _{A\subset \mathbb{Z}} |P(A)-Q(A)|.
\]
\end{definition}

We use the following result from \cite{ehm91}.

\begin{lemma}[\cite{ehm91}]
\label{lem:total-variation}
For $\Hyp(n,m,k)$ and $\Bin(k,p)$ where $p=\frac{m}{n}$ and $kp(1-p)\geq 1$,
\[
\delta(\Hyp(n,m,k),\Bin(k,p))\leq \frac{k-1}{n-1}.
\]
\end{lemma}

\section{Tight Bounds for Partition Matroids}\label{sec:partitionFull}

Recall the definition of partition matroids.

\begin{definition}[Partition Matroid] A partition matroid $\M=(E,\I)$ is defined by a ground set $E$ being partitioned into disjoint sets $A_1,\dots,A_m$ and $m$ integers $b_1,\dots,b_m$ where $0\leq b_i\leq |A_i|$. A set $S\subseteq E$ is independent iff $|S\cap A_i|\leq b_i$ for every $0\leq i\leq m$. We refer to $A_i$ as a \emph{part}, and $b_i$ as the \emph{budget} of the part.
\end{definition}

In this section, we present an $O(n^{1/3}\log n)$ round algorithms for finding bases in partition matroids, which matches the lower bound $\widetilde\Omega(n^{1/3})$ of \cite{kuw85} up to logarithmic factors.

\subsection{A Randomized Algorithm}

In a partition matroid $\M=(E,\I)$ with $|E|=n$, recovering a single part in each round is straightforward, but very inefficient as there can be $\Omega(n)$ parts. The main idea of our algorithm is to recover multiple parts simultaneously in a single round. Specifically, if we consider adding elements in a random order until the budget of some part is exceeded, we will show that we can efficiently identify and recover this part. If we repeat this $\text{poly}(n)$ times in parallel, we can collect and identify all the parts that ever caused a dependency and remove them from the matroid. Intuitively, in the next iteration, we expect that the dependence will occur later when adding elements in random order, as the parts that are more likely to cause circuits have already been removed (i.e., the small budget parts). Thus, the expected number of elements needed to observe a dependency increases. We formally quantify this growth and thereby show that in $\tilde O(n^{1/3})$ rounds, we can recover all parts in the matroid.

We begin by presenting a simple algorithm to recover a single part. The algorithm adds elements according to the order of a permutation $\pi$, and stops once a dependence occurs. This implies that for exactly one part, the budget has been exceeded by 1, and allows us to uniquely identify this part.

\begin{algorithm}[ht]
    \caption{RecoverSinglePart($\M=(E,\I),\pi: \text{bijection }[n]\to E)$)}
    \label{alg:recover-single-part}
    \For {$i \in [n]$ in parallel} {
        Query $\Ind(\{\pi(1),\dots,\pi(i)\})$
    }
    Let $t$ be the smallest index such that $S=\{\pi(1),\dots,\pi(t-1)\}\in \I$ and $S+\pi(t)\not\in \I$\\ \label{line:definition-t}
    $I \gets \emptyset$\\
	\For {$i=1,\dots,t-1$} {
        Query $\Ind(S-\pi(i)+\pi(t))$\\
        \If{$\Ind(S-\pi(i)+\pi(t))=1$} {
            $I \gets I\cup \pi(i)$
        }
    }
    $T \gets I\cup \pi(t)$\\ \label{line:find-circuit}
    \For {$i=t+1,\dots,n$} {
        Query $\Ind(I+\pi(i))$
    }
    \If{$\Ind(I+\pi(i)) = 1$} {
        $T\gets A\cup \pi(i)$
    }
    \Return $T$ and $|I|$
\end{algorithm}

\begin{claim}
\label{lem:recover-single-part}
In a partition matroid $\M$ where every part has budget $\geq 1$, \cref{alg:recover-single-part} finds the part that contains $\pi(t)$ ($t$ as defined in \cref{line:definition-t}), and can be implemented in $2$ rounds.
\end{claim}

\begin{proof}
Let $A_\ell$ be the part that contains $\pi(t)$. Given that $\Ind(S) = 1$ and $\Ind(S+\pi(t))=0$, it follows that $|S\cap A_{\ell}|=b_{\ell}$ and $|S\cap A_j|\leq b_j$ for every $j\neq \ell$. Therefore, for every $i<t$, $S-\pi(i)+\pi(t)$ is independent if and only if $\pi(i)\in A_{\ell}$. As $|S\cap A_{\ell}| = b_{\ell}$, we have $I\subseteq A_{\ell}$ and $|I|=b_{\ell}$. Moreover, for every $i>t$, we see that $I+\pi(i)$ is dependent if and only if $\pi(i)\in A_{\ell}$. Thus, we conclude that $T=A_{\ell}$ and $|I|=b_{\ell}$.

Note that the only adaptivity we need is to find $I$. The for loops before \cref{line:find-circuit} can be implemented in 1 round in parallel (by querying all prefixes and prefixes with one element excluded) 
, and the for loop after \cref{line:find-circuit} can also be implemented in 1 round.
\end{proof}

We also show a similar subroutine to recover every part with budget at most $50$.

\begin{claim}
\cref{alg:remove-small-parts} eliminates all parts with budget at most 50, and can be implemented in 1 round.
\end{claim}

\begin{algorithm}[ht]
    \caption{RemoveSmallParts($\M=(E,\I)$)}
    \label{alg:remove-small-parts}
    \For{$i\in [50], S\in {E\choose i}$ in parallel} {
        Query $\Ind(S)$
    }
    $B\gets \emptyset$\\
    \For{$i\in[49]$} {
        \For{$S\in {E\choose i},x\in E\setminus S$}{
            \If{$\Ind(S)=1\land\Ind(S+x)=0$} {
                $T\gets S$
                \For{$y\in E\setminus S$} {
                    \If{$\Ind(S+y)=1$} {
                        $T\gets T+ y$
                    }
                }
                $B\gets B\cup S$\\
                $\M\gets \M\setminus T$
            }
        }
    }
    \Return $\M,B$
\end{algorithm}

Now we are ready to present our main procedure, detailed in \cref{alg:recover-multiple-parts}. It begins by invoking \cref{alg:remove-small-parts} to remove parts with small budgets. This is for a technical reason in the analysis of our algorithm. It also checks the trivial case that the whole ground set is independent. Then, the algorithm draws $\text{poly}(n_0)$ random permutations $\pi$ in parallel and checks if the first $n/1000$ elements of $\pi$ form an independent set. If the set is independent, the algorithm adds it to the solution, contracts it from the matroid, and then terminates. Otherwise, we invoke \cref{alg:recover-single-part} to recover the part which contains the first circuit that appears in the matroid when elements are added in the order of $\pi$. The recovered parts are recorded and removed from the matroid at the end of each iteration.

\begin{algorithm}[ht]
    \caption{RecoverMultipleParts($\M=(E,\I)$)}
    \label{alg:recover-multiple-parts}
    $\M,B \gets \text{RemoveSmallParts}(\M)$\\
    \If{$\Ind(E)=1$} {
        \Return $\emptyset, E$
    }
    $k\gets 0$\\
    \While{$\M \neq \emptyset$} {
        $\mathcal{A}\gets \emptyset, I\gets \emptyset$\\
        \For {$i\in [n_0^{10}]$ in parallel} { \label{line:for-loop}
            Draw a random permutation (bijection) $\pi:[n]\to E$\\
            Query $\Ind(\{\pi(1),\dots,\pi(n/1000)\}$\\
            \If{$\Ind(\{\pi(1),\dots,\pi(n/1000)\})= 1$} {
                $I\gets \{\pi(1),\dots,\pi(n/1000)\}$
            }
            $T,\ell\gets \text{RecoverSinglePart}(\M, \pi)$\\
            $\mathcal{A}\gets \mathcal{A}\cup \{(T,\ell)\}$
        }
        \If{$I\neq \emptyset$} {
            \Return $\M/I, B\cup I$ \label{line:return}
        }
        $k\gets k+1, S_k\gets\emptyset$\\
        \For{$(T,\ell)\in \mathcal{A}$} {
            $S_k\gets S_k\cup T$\\
            Pick an arbitrary $I\in {T\choose \ell}$, $B \gets B\cup I$
        }
        $\M \gets \M \setminus S_k$
    }
    \Return $B$
\end{algorithm}

Let $S_1,\dots,S_k$ be the sets recovered by \cref{alg:recover-multiple-parts} in $k$ iterations. Note that each $S_i$ is the union of one or more parts. For every $i\in [k]$, we define $\alpha(S_i)$ as the smallest integer $\ell$ such that
\[
\Pr_{T\in {S_i\choose\ell}} [\Ind(T)=1]\leq \frac{1}{2}.
\]
I.e., the probability that a random subset of $S_i$ of cardinality $\ell$ is independent is less that $1/2$. It follows that within $S_i$, a random subset of cardinality less than $\alpha(S_i)$ is independent with probability at least $1/2$, and a random subset of cardinality at least $\alpha(S_i)$ is dependent with probability at least $1/2$.

We analyze the performance of \cref{alg:recover-multiple-parts} in the following claims. Note that the algorithm requires $O(k)$ rounds of adaptivity, as all queries are made within the for loop on \cref{line:for-loop} and can be executed in parallel, where each parallel instance requires $O(1)$ rounds by \cref{lem:recover-single-part}. Thus, our ultimate goal is to bound $k=\tilde O(n^{1/3})$.

\begin{claim}
For every $i\in[k]$, $\alpha(S_i)\geq 50$.
\end{claim}
\begin{proof}
Recall that we invoke \cref{alg:remove-small-parts} to remove all parts with budget less than or equal to $50$. Since in a partition matroid, the size of a dependent set must be at least the budget of some part, we see that $\alpha(S_i)\geq 50$ according to the definition of $\alpha$.
\end{proof}

\begin{claim}
\label{lem:estimation}
In the $i$-th iteration of \cref{alg:recover-multiple-parts}, let $A_1,\dots,A_{\ell}$ be the remaining parts. For every $j\in [\ell]$, let $p_j$ denote the probability that \cref{alg:recover-single-part}, when given a random permutation, returns $A_j$. Then with probability at least $1-2^{-n_0^7}$, we have for any $j\in[\ell]$ with $p_j\geq 1/n^2$, $A_j\subseteq S_i$.
\end{claim}
\begin{proof}
Consider any $j\in[\ell]$ with $p_j\geq 1/n^2$, the probability that the algorithm fails to find it in the $n_0^{10}$ parallel instances of \cref{alg:recover-single-part} is at most
\[
\left(1-\frac{1}{n^2}\right)^{n_0^{10}} \leq e^{-n_0^8}.
\]
Taking a union bound over at most $\ell\leq n\leq n_0$ parts, we see that the algorithm fails to find any such part with probability at most $e^{-n_0^8}\cdot n_0\leq 2^{-n_0^7}$.
\end{proof}

It immediately follows from the above claim that with probability at least $1-1/n$, the first circuit will appear inside $S_i$ when adding elements according to the order of a random permutation $\pi$.

\begin{claim}
\label{lem:alpha-upper-bound}
In the $i$-th iteration of \cref{alg:recover-multiple-parts}, suppose with probability at least $1-1/n$, the first circuit appears inside $S_i$ when adding elements according to the order of a random permutation $\pi$. If
\[
\frac{\alpha(S_i)}{|S_i|} \geq \frac{1}{250},
\]
then the algorithm will terminate on \cref{line:return} in this iteration with probability at least $1-2^{-n_0^{9}}$.
\end{claim}
\begin{proof}
Let $\ell = \frac{\alpha(S_i)}{4|S_i|}n\geq n/1000$. Suppose $U$ is a random subset of $E$ of cardinality $\ell$, we define random variable $X_i=|U\cap S_i|$. Note that $X_i\sim \Hyp(n,|S_i|,\ell)$, and $\E[X_i]=\alpha(S_i)/4$. By Markov's inequality,
\[
\Pr[X_i\geq \alpha(S_i)] \leq \frac{1}{4}.
\]
Since $\ell \geq n/1000$, we see that
\begin{align*}
\Pr_\pi[\Ind(\{\pi(1),\dots,\pi(n/1000)\})=0] & \leq \Pr_\pi[\Ind(\{\pi(1),\dots,\pi(\ell)\})= 0] \\
& \leq \frac{1}{n}+\Pr_U[\Ind(U\cap S_i)=0]\\
& \leq \frac{1}{n}+\left(\Pr_{T\in {S_i\choose X_i}}[\Ind(T)=0\mid X_i< \alpha(S_i)] + \Pr[X_i\geq \alpha(S_i)]\right)\\
& \leq \frac{1}{n}+\frac{1}{2}+\frac{1}{4} \leq \frac{7}{8}.    
\end{align*}

Once the first $n/1000$ elements of $\pi$ form an independent set, the algorithm will contract the set and terminate on \cref{line:return}. Therefore, the probability that the algorithm does not terminate is at most $(7/8)^{n_0^{10}} \leq 2^{-n_0^9}$.
\end{proof}

By taking a simple union bound, the algorithm achieves all the above guarantees with probability at least $1-2^{n_0^7}-2_{n_0}^9\geq 1-2^{n_0^6}$. We make subsequent claims conditioned on this event.

\begin{claim}
\label{lem:alpha-increse}
Let $S_1, \dots S_k$ be the sets obtained by \cref{alg:recover-multiple-parts}. For any $i<j$ where $|S_i|\leq 2|S_j|$ and $|S_i|,|S_j|\leq n^{2/3}$, we have
\[
\alpha(S_j) \geq \frac{\alpha(S_i)|S_j|}{|S_i|} + \frac{1}{5\sqrt{2}}\sqrt{\frac{\alpha(S_i)|S_j|}{|S_i|}}.
\]
\end{claim}

\begin{proof}
Let $\ell = \frac{\alpha(S_i)}{|S_i|} n$. Suppose $U$ is a subset of the ground set $E$ of cardinality $\ell$, we define random variables $X_i = |U\cap S_i|, X_j =|U \cap S_j|$. Note that $X_i\sim \Hyp(n,|S_i|,\ell),X_j\sim \Hyp(n,|S_j|,\ell)$ and they are negatively correlated.

For the sake of contradiction, suppose
\[
\alpha(S_j) \leq \frac{\alpha(S_i)|S_j|}{|S_i|} +  \frac{1}{5\sqrt{2}} \sqrt{\frac{\alpha(S_i)|S_j|}{|S_i|}}.
\]
We aim to show that
\[
\Pr[X_i< \alpha(S_i) \land X_j\geq \alpha(S_j)] = \Omega(1).
\]
Consider the $i$-th iteration of \cref{alg:recover-multiple-parts}, the above implies that for a random permutation $\pi$ of $[n]$, there is a constant probability that there are less than $\alpha(S_i)$ elements from $S_i$ and more than $\alpha(S_j)$ elements from $S_j$ in the first $\ell$ elements of $\pi$. Conditioned on this, with at least $1/4$ probability, there is no circuit within $S_i$ but there is a circuit in $S_j$. This implies that with $\Omega(1)$ probability, the first circuit appears outside of $S_i$ when adding elements according to the order of a random permutation $\pi$, which contradicts \cref{lem:estimation}.

To estimate $X_i\sim \Hyp(n,|S_i|,\ell)$, we define $Y_i\sim \Bin(\ell,|S_i|/n)$. It gives a good estimation since $\ell/n = \alpha(S_i)/|S_i|\leq 1/250$ by \cref{lem:alpha-upper-bound}. Since $|S_i|\leq n^{2/3}$, we see that
\[
\E[Y_i] = \alpha(S_i),\quad \Var[Y_i] = \ell \frac{|S_i|}{n}\left(1-\frac{|S_i|}{n}\right) \geq  \frac{\alpha(S_i)}{2}\geq 25
\]
It follows from \cref{lem:binomial} that
\[
\Pr[Y_i<\alpha(S_i)] = \Pr[Y_i\leq \E[Y_i]-1] \geq \Pr\left[Y_i\leq \E[Y_i]-\frac{1}{5}\sqrt{\Var[Y_i]}\right]\geq \frac{1}{108}.
\]
Combined with \cref{lem:total-variation} and $\ell/n\leq 1/250$, we conclude that
\[
\Pr[X_i<\alpha(S_i)] \geq \frac{1}{108}-\frac{\ell-1}{n-1}\geq \frac{1}{200}.
\]

To estimate $X_j\sim \Hyp(n,|S_j|,\ell)$, we also define $Y_j\sim \Bin(\ell,|S_j|/n)$. Since $|S_j|\leq n^{2/3}, |S_i|\leq 2|S_j|$, we see that
\[
\E[Y_j]=\frac{\alpha(S_i)|S_j|}{|S_i|},\quad \Var[Y_j]=\ell \frac{|S_j|}{n}\left(1-\frac{|S_j|}{n}\right) \geq  \frac{\alpha(S_i)|S_j|}{2|S_i|}\geq \frac{\alpha(S_i)}{4} \geq 1.
\]
It follows from \cref{lem:binomial} that
\[
\Pr\left[Y_j\geq \E[Y_j]+\frac{1}{5}\sqrt{\Var[Y_j]}\right] \geq \frac{1}{108}.
\]
Combined with \cref{lem:total-variation} and $\ell/n\leq 1/250$, we have
\[
\Pr\left[X_j\geq \E[Y_j]+\frac{1}{5}\sqrt{\Var[Y_j]}\right] \geq \frac{1}{108} - \frac{\ell-1}{n-1} \geq \frac{1}{200}.
\]
We conclude that
\[
\Pr[X_j\geq \alpha(S_j)]\geq \Pr\left[X_j\geq \frac{\alpha(S_i)|S_j|}{|S_i|}+\frac{1}{5\sqrt{2}}\sqrt{\frac{\alpha(S_i)|S_j|}{|S_i|}}\right] \geq \Pr\left[X_j\geq \E[Y_j]+\frac{1}{2}\sqrt{\Var[Y_j]}\right] \geq \frac{1}{200}.
\]
The penultimate inequality follows from
\[
\E[Y_j] = \frac{\alpha(S_i)|S_j|}{|S_j|},\quad \Var[Y_j]\geq\frac{\alpha(S_i)|S_j|}{2|S_i|}. 
\]

Since $X_i,X_j$ are negatively correlated, we have
\[
\Pr[X_i<\alpha(S_i)\land X_j\geq \alpha(S_j)] \geq \Pr[X_i<\alpha(S_i)] \cdot \Pr[X_j\geq \alpha(S_j)] \geq \frac{1}{200^2}.
\]
as desired.
\end{proof}

\begin{lemma}
\label{lem:round-complexity}
Throughout the course of \cref{alg:recover-multiple-parts}, $k = O(n^{1/3})$.
\end{lemma}

\begin{proof}
Consider $S_1,\dots,S_k$ obtained from \cref{alg:recover-multiple-parts}. We first note that there are at most $n^{1/3}$ $S_i$'s with cardinality at least $n^{2/3}$. Therefore, we will drop every set of cardinality larger that $n^{2/3}$ and assume that $|S_i|\leq n^{2/3}$ for every $i\in[k]$.

For any $\ell\in [2/3 \cdot\log n]$, let $S_{i_1},\dots,S_{i_{q(\ell)}}$ be all the $S_i$'s with $|S_{i}|\in[2^{\ell-1},2^{\ell}-1]$ and assume $i_1<\dots<i_{q(\ell)}$. As $|S_{i_{j}}|\leq 2|S_{i_{j+1}}|$ and $|S_{i_j}|,|S_{i_{j+1}}|\leq n^{2/3}$, it follows from \cref{lem:alpha-increse} that
\[
\frac{\alpha(S_{i_{j+1}})}{|S_{i_{j+1}}|} \geq \frac{\alpha(S_{i_j})}{|S_{i_j}|} + \frac{1}{5\sqrt{2}} \sqrt{\frac{\alpha(S_{i_j})}{|S_{i_j}||S_{i_{j+1}}|}} \geq \frac{\alpha(S_{i_j})}{|S_{i_j}|} + \frac{1}{5\sqrt{2} \cdot 2^{\ell/2}} \sqrt{\frac{\alpha(S_{i_j})}{|S_{i_j}|}}.
\]
Since $\alpha(S_{i_1})/|S_{i_1}|\geq 1/n, \alpha(S_{q(\ell)})/|S_{q(\ell)}|\leq 1$, the recursion gives $q(\ell)=O(2^{\ell/2})$. Therefore,
\[
k\leq \sum_{\ell =0}^{\frac{2}{3}\log n} q(\ell) = \sum_{\ell=0}^{\frac{2}{3}\log n}O(2^{\ell/2}) = O(n^{1/3}).
\]
This concludes the proof.
\end{proof}

We obtain our algorithm for finding a basis by iteratively executing \cref{alg:recover-multiple-parts}.

\begin{algorithm}
    \caption{FindBasis($\M=(E,\I)$)}
    \label{alg:find-basis}
    $B\gets \emptyset$\\
    \While{$\M \neq \emptyset$} {
        $\M,I\gets \text{RecoverMultipleParts}(\M)$\\
        $B\gets B\cup I$
    }
    \Return $B$
\end{algorithm}

\begin{theorem}
For a partition matroid $\M=(E,\I)$ with $|E|=n$, \cref{alg:find-basis} requires $O(n^{1/3}\log n)$ rounds to recover a basis in $\M$ with high probability.
\end{theorem}

\begin{proof}
Throughout the algorithm, we either contract an independent set of size $\Omega(n)$ or recover parts. When a part is recovered, an independent set of size equal to its budget is always included before removing the part from the matroid. Therefore, the final set obtained is indeed a basis of the matroid.

In each execution of \cref{alg:recover-multiple-parts}, it either finds an independent set and terminates at \cref{line:return}, or it successfully recovers the entire matroid. If it terminates at \cref{line:return}, it implies that an independent set of size $n/1000$ has been found and the matroid has been contracted. It is straightforward to check that the matroid remains a partition matroid after contraction. Since this can happen $O(\log n)$ times, \cref{alg:recover-multiple-parts} is executed at most $O(\log n)$ times. Combined with \cref{lem:round-complexity}, the total rounds of adaptivity required is $O(n^{1/3}\log n)$. By a simple union bound, we can also bound the total failure probability by $2^{-\Omega(n)}$.
\end{proof}

\subsection{Derandomization}
We also show that the above algorithm can be derandomized.

\begin{claim}
\label{lem:derandomize-permutation}
There exists a universal family of permutations $\pi_1,\dots,\pi_{n_0^{10}}$ such that for any partition matroid $\M=(E,\I)$ with $|E|=n\leq n_0$, the guarantees of \cref{lem:estimation} and \cref{lem:alpha-upper-bound} are satisfied after running \cref{alg:recover-multiple-parts} using these permutations instead of random permutations.
\end{claim}
\begin{proof}
By \cref{lem:estimation} and \cref{lem:alpha-upper-bound}, we see that a random collection of $\pi_1,\dots,\pi_{n_0^{10}}$ achieves the guarantees with probability at least $1-2^{n_0^6}$. Since there are at most $n^n \cdot n!\leq 2^{n^2}$ possible partition matroids, we obtain that a random collection of $\pi_1,\dots,\pi_{n_0^{10}}$ achieves the guarantees for any partition matroid with probability at least $1-2^{-n_0^6}\cdot2^{n^2}\geq 1-2^{-n_0^5}$. This implies that there exists a deterministic choice of $\pi_1,\dots,\pi_{n_0^{10}}$ that achieves the guarantees.
\end{proof}

\begin{theorem}
There is a {\em deterministic} algorithm that finds a basis of any partition matroid $\M$ in $\tilde{O}(n^{1/3})$ adaptive rounds, using only polynomially many independence queries per round.
\end{theorem}
\begin{proof}
For every $i\in [n]$, by \cref{lem:derandomize-permutation}, there exists a feasible family of permutations $\mathcal{P}_i=\{\pi_1,\dots,\pi_{n^{10}}\}$ for matroids on $i$ elements, which we encode non-uniformly. The remainder of the algorithm is identical to \cref{alg:find-basis}, except that we replace the random permutations in \cref{alg:recover-multiple-parts} with the deterministic family of permutations obtained earlier.
\end{proof}

Note that this derandomization is non-uniform. While such derandomizations are typical in some settings, such as $\textbf{BPP} \subseteq \textbf{P / poly}$, non-uniform derandomizations are not always possible with query complexity bounds, where randomization can sometimes be essential. 

\section{Decomposition Algorithm}\label{sec:decomp}

In this section, we present our decomposition algorithm, and derive all of the relevant results that we will need when designing our improved algorithm for finding bases. 

\subsection{Finding Sets with Many Circuits}

To start, we present an algorithm which adds elements in the order of a permutation $\pi$ until a circuit forms:

\begin{algorithm}
    \caption{FindCircuit$(\M=(E,\I),\pi: \text{bijection }[n]\to E)$}\label{alg:find-circuit}
    \For {$i \in [n]$ in parallel} {
        Query $\Ind(\{\pi(1),\dots,\pi(i)\})$
    }
    Let $t$ be the smallest index such that $S=\{\pi(1),\dots,\pi(t-1)\}\in \I$ and $S+\pi(t)\not\in \I$. \\
    $C_\pi\gets \{\pi(t)\}$ \\
    \For {$i=1,\dots,t-1$ in parallel}{
            Query $\Ind(S-\pi(i)+\pi(t))$. \\
            \If{$\Ind(S-\pi(i)+\pi(t))=1$}{
                $C_\pi \gets C_\pi + \pi(i)$
            }
        }
    \Return{$C_\pi$}
\end{algorithm}

The following claim follows immediately from \cref{lem:matroid-circuit}.

\begin{claim}
\cref{alg:find-circuit} can be implemented in 1 round and returns the first circuit appears when adding elements in the order of a permutation $\pi$. 
\end{claim}

Using the above procedure, we define the following quantities:

\begin{definition}
For a matroid $\M=(E,\I)$ and an element $x\in E$, we let
\[
    p_{\M}(x) = \Pr_{\pi}[x\in\mathrm{FindCircuit}(\M,\pi)]
\]
Similarly, for a subset $S \subseteq E$, we let
\[
    q_{\M}(S) = \Pr_{\pi}[\mathrm{FindCircuit}(\M,\pi)\subseteq S]
\]
We let $\widehat{q}_\M(S)$ denote the estimate of this probability that results from running \cref{alg:find-circuit} on $n_0^{10}$ random permutations. 
\end{definition}

\begin{remark}
We sometimes omit the subscript $\M$ when the underlying matroid is clear from the context.
\end{remark}

A simple application of a Chernoff bound yields the following statement:

\begin{claim}\label{clm:boundedErrorEstimation}
    For a matroid $\M=(E,\I)$ and every subset $S \subseteq E$, 
    \[
    |\widehat{q}(S) - q(S)| \leq \frac{1}{n_0^2},
    \]
    with probability $1 - 2^{-n_0}$.
\end{claim}

\begin{remark}
    Note that because the error probability is $1 - 2^{-n_0}$, we will often present intermediate claim / theorem statements without quantifying their success probability. Ultimately, our algorithm will only ever perform $\mathrm{poly}(n_0)$ invocations of the decomposition, and thus this error probability is negligible. 
\end{remark}

\begin{definition}
For a matroid $\M=(E,\I)$ and $S\subseteq E$, we define $\alpha(S)$ as the smallest integer $\ell$ such that
\[
\Pr_{T\in {S\choose\ell}} [\Ind(T)=1]\leq \frac{1}{2}.
\]
I.e., the probability that a random subset of $S$ of cardinality $\ell$ is independent is less that $1/2$. Equivalently, $\alpha(S)$ is the median number of elements required until a circuit appears when running $\text{FindCircuit}(\M|_S,\pi)$ on a random permutation $\pi$. We also use $\widehat \alpha(S)$ to denote the estimate of $\alpha(S)$ results from running \cref{alg:find-circuit} on $n_0^{10}$ random permutations and taking median.
\end{definition}

We have the following useful properties.

\begin{claim} \label{clm:alpha-dependent}
For any set $S$ and any integer $d>1$, a random subset of $S$ of cardinality $d\cdot\alpha(S)$ is dependent with probability at least $1-2^{-d}$.
\end{claim}
\begin{proof}
Let $T$ be a random subset of $S$ of cardinality $d\cdot\alpha(S)$, and $R_1,\dots,R_{d}$ be independently drawn random subsets of $S$ of cardinality $\alpha(S)$. We see that
\[
\Pr[\Ind(T)=1] \leq \Pr\left[\Ind\left(\bigcup_{i\in [d]} R_i\right)=1\right]\leq \prod_{i=1}^{d} \Pr[\Ind(R_i)=1] \leq (1/2)^{d}.
\]
The last inquaility follows from the definition of $\alpha(S)$.
\end{proof}

\begin{claim} \label{clm:alpha-estimation}
$(\alpha(S)-1)/2\leq \widehat \alpha(S)\leq 2\cdot \alpha(S)$ with probability at least $1-2^{-n_0}$.
\end{claim}
\begin{proof}
By \cref{clm:alpha-dependent}, the probability that a random subset of $S$ of cardinality $2\cdot\alpha(S)$ is independent is at most $1/4$. Thus, we see that $\widehat \alpha(S)\leq 2\cdot\alpha(S)$ with high probability by a straightforward Chernoff bound. On the other hand, we show that the probability a random subset of $S$ of cardinality $(\alpha(S)-1)/2$ is independent is at least $1/\sqrt{2}$. This is due to an argument similar to \cref{clm:alpha-dependent}: let $T$ be a random subset of $S$ of cardinality $\alpha(S)-1$ and $R_1,R_2$ be 2 independently drawn random subsets of $S$ of cardinality $(\alpha(S)-1)/2$. We have
\[
\frac{1}{2}\leq \Pr[\Ind(T)=1] \leq \Pr[\Ind(R_1\cup R_2)=1] \leq \Pr[\Ind(R_1)=1]^2,
\]
and thus $\Pr[\Ind(R_1)=1]\geq 1/\sqrt{2}$. Again, we have $\widehat \alpha(S)\geq (\alpha(S)-1)/2$ with high probability by a Chernoff bound.
\end{proof}

Now, we will let $S^* \subseteq E$ be a \emph{greedily-optimal} set in the following sense:

\begin{definition}
    For a matroid $\M=(E,\I)$, we say that a set $S^* \subseteq E$ is \emph{greedily-optimal} if 
    \[
    \widehat{q}(S^*) \geq 1-2^{-20} + \frac{\sum_{i = 1}^{|S^*|} \frac{1}{i}}{2^{20}\log n},
    \]
    and there is no element $x \in S^*$ such that 
    \[
    \widehat{q}(S^* - x) \geq 1-2^{-20} + \frac{\sum_{i = 1}^{|S^*|-1} \frac{1}{i}}{2^{20}\log n}.
    \]
\end{definition}

Observe that there is a simple algorithm for creating greedily-optimal sets: we start with the complete set $n$, and continue to delete elements until the condition no longer holds:

\begin{algorithm}[H]
\caption{FindGreedilyOptimal$(\M=(E,\I))$}\label{alg:FindGreedilyOptimal}
Multiset $\mathcal{C}\gets \emptyset$\\
\For {$i\in[n_0^{10}]$ in parallel} {
    Draw a random permutation (bijection) $\pi:[n]\to E$\\
    $C_\pi\gets\text{FindCircuit}(\M,\pi)$\\
    $\mathcal{C}\gets \mathcal{C}\cup \{C_\pi\}$
}
$S^* \gets E$. \\
\While{True}{
\For{$x \in S^*$}{
$\widehat{q}(S^* - x) 
\gets\frac{ \left |C\in \mathcal{C}: C \subseteq S^* - \{x \} \right |}{|\mathcal{C}|}$
}

\If{$\exists x \in S^*$ s.t. $ \widehat{q}(S^* - x) \geq 1-2^{-20} + \frac{\sum_{i = 1}^{|S^*|-1} \frac{1}{i}}{2^{20}\log n}$} {
Let $x^*$ be the first such element\\
$S^* \leftarrow S^* - x^* $
} 
\Else {
\Return{$S^*$}
}
}
\end{algorithm}

\begin{claim}
    \cref{alg:FindGreedilyOptimal} finds a greedily-optimal set $S^*$ in $\M$. 
\end{claim}

\begin{proof}
Note that at the initialization of \cref{alg:FindGreedilyOptimal}, $S^* = E$, and so it must be the case that $1 = \widehat{q}(S^*) = \widehat{q}(E) \geq 1-2^{-20} + \frac{\sum_{i = 1}^{n} \frac{1}{i}}{2^{20}\log n}$, as 
    \[
    1-2^{-20} + \frac{\sum_{i = 1}^{n} \frac{1}{i}}{2^{20}\log n} \leq 1-2^{-20} + \frac{1 + \ln n}{2^{20} \log n} \leq 1-2^{-20} + \frac{1}{2^{20}\log n} + \frac{\log n}{2^{20}\log(e)} < 1.
    \]
    Now, in each iteration of \cref{alg:FindGreedilyOptimal}, we only remove elements $x$ from $S^*$ that ensure that $S^*$ continues to satisfy 
    \[
    \widehat{q}(S^*) \geq 1-2^{-20} + \frac{\sum_{i = 1}^{|S^*|} \frac{1}{i}}{2^{20}\log n}.
    \]
    At the termination of the above algorithm, we are also guaranteed that there is no $x \in S^*$ for which 
    \[
    \widehat{q}(S^* - x)\geq 1-2^{-20}+ \frac{\sum_{i = 1}^{|S^*|-1} \frac{1}{i}}{2^{20}\log n},
    \]
    thereby yielding our claim. 
\end{proof}

Now, we establish the following claim, which seeks to understand the \emph{marginal} probabilities that an element $x \in S^*$ participates in a circuit.

\begin{claim}\label{clm:highProbCircuit}
Let $\M=(E,\I)$ be a matroid on $n$ elements, and let $S^*$ be a greedily-optimal set. Let $\M' = \M|_{S^*}$ be the matroid restricted to $S^*$. Then, for every $x \in S^*$,
\[
    p_{\M'}(x) \geq \frac{1}{2\cdot 2^{20} |S^*| \log n}.
\]
\end{claim}

\begin{proof}
    First, observe that by definition of being greedily-optimal, for every element $x \in S^*$, it must be that 
    \[
    \widehat{q}_{\M}(S^* - x) < 1- 2^{-20} + \frac{\sum_{i = 1}^{|S^*|-1} \frac{1}{i}}{2^{20}\log n},
    \]
    and that 
    \[
    \widehat{q}_{\M}(S^*) \geq 1- 2^{-20}+ \frac{\sum_{i = 1}^{|S^*|} \frac{1}{i}}{2^{20}\log n}.
    \]
    In particular, this means
    \[
    \widehat{q}_{\M}(S^*)  - \widehat{q}_{\M}(S^* - x) \geq \frac{1}{2^{20}|S^*|\log n}.    \]
    By our bound relating $p$ and $\widehat{p}$ (\cref{clm:boundedErrorEstimation}), we also know that with overwhelmingly high probability,
    \[
    q_{\M}(S^*) - q_{\M}(S^* - x)\geq \frac{1}{2\cdot2^{20}|S^*|\log n}.
    \]

    Now, let us define some auxiliary values: for a matroid $\M=(E,\I)$ and $x\in T, T\subseteq E$, $p_{\M}(x,T)$ is defined as
    \[
    p_{\M}(x,T) = \Pr_{\pi}[\{x\}\subseteq \mathrm{FindCircuit}(\M,\pi)\subseteq T]
    \]
    where the permutation $\pi$ is drawn uniformly at random. We can observe that $p_{\M'}(x) = p_{\M'}(x,S^*) \geq p_{\M}(x,S^*)$. The inequality is because whenever we sample in accordance to a permutation $\pi$ of $[n]$, and recover a circuit $C_{\pi}$ such that $x \in C_{\pi}$ and $C_{\pi} \subseteq S^*$, the same permutation, if restricted to $S^*$ and used to sample elements of $S^*$, would have given a circuit such that $x \in S^*$.

    Finally, we can observe that $p_\M(x,S^*) = q_{\M}(S^*) - q_{\M}(S^* - x)$, as $q_{\M}(S^*) - q_{\M}(S^* - x)$ is exactly the probability that a circuit, when sampled from $\M$, is contained in $S^*$ and uses the element $x$. Thus, we conclude that $p_{\M'}(x) \geq p_\M(x,S^*) \geq \frac{1}{2\cdot2^{20}|S^*|\log n}$.
\end{proof}

\subsection{Iterative Matroid Decomposition}

Now that we have established how to recover greedily-optimal sets, we show how we can repeat this procedure to iteratively decompose our starting matroid. To start, we remove all circuits of length $\leq 50$ to ensure that our algorithm has a non-trivial starting point. 

\begin{algorithm}
    \caption{RemoveSmallCircuits$(\M=(E,\I))$}\label{alg:RemoveSmallCircuits}
    \For{$i\in [50], S\in {E\choose i}$ in parallel} {
        Query $\Ind(S)$.
    }
    Fix an arbitrary bijection $\pi:E\to [n]$\\
    \For{$i\in [50], S\in {E\choose i}$} {
        $x \gets \arg\min_{y\in S} \pi(y)$\\
        \If{$\Ind(S-x)=1\land \Ind(S)=0$} {
            $\M\gets \M\setminus \{x\}$.
        }
    }
    \Return{$\M$}
\end{algorithm}

\begin{claim}
\cref{alg:RemoveSmallCircuits} can be implemented in 1 round and removes all circuits of size $\leq 50$ in $\M$ while ensuring the rank of $\M$ remains unchanged.
\end{claim}
\begin{proof}
It is clear that the algorithm finds all circuits of size $\leq 50$. Since it removes at least 1 element from each such circuit, these circuits are indeed eliminated. We show the rank of the matroid remains unchanged in the following.

Let $E$ be the ground set of the input matroid and $S$ be the set of elements we deleted from $\M$ during \cref{alg:RemoveSmallCircuits}. We denote the elements in $S$ as $e_1,\dots, e_{|S|}$ and order them such that $\pi(e_1)<\dots<\pi(e_{|S|})$. We prove $\rank(E)=\rank(E\setminus S)$ by induction on the size of $S$: Suppose $\rank(E)=\rank\left(E\setminus \bigcup_{j<i} \{e_j\}\right)$, we see that $e_i$ must be in a circuit $C$ which is disjoint from $\bigcup_{j<i} \{e_j\}$ since we always delete the smallest element w.r.t. $\pi$. Thus, we have
\[
e_i \in \text{span} \left(C\setminus \{e_{i}\}\right) \subseteq \text{span}\left(\left(E\setminus \bigcup_{j< i} \{e_j\}\right)\setminus\{e_i\}\right) = \text{span} \left(E\setminus \bigcup_{j\leq i} \{e_j\}\right)
\]
and
\[
\rank\left(E\setminus \bigcup_{j\leq i} \{e_j\}\right) = \rank\left(E\setminus \bigcup_{j< i} \{e_j\}\right) = \rank(E).
\]
\end{proof}

Having removed all short circuits, we next consider  repeatedly running \cref{alg:FindGreedilyOptimal}, peeling off sets $S_1, S_2, \dots $:

\begin{algorithm}
    \caption{Peel$(\M)$}\label{alg:peel}
    $S \leftarrow \mathrm{FindGreedilyOptimal}(\M)$. \\
    $\M \leftarrow \M \setminus S$  \\
    \Return{$S, \M$}
\end{algorithm}

\begin{algorithm}[H]
    \caption{IterativePeel$(\M)$}\label{alg:iterativePeel}
    $\M\gets \text{RemoveSmallCircuits}(\M)$\\
    $k = 0$. \\
    \While{$\M \neq \emptyset$}{
    $k \leftarrow k + 1$.\\
    $S_k, \M \leftarrow \mathrm{Peel}(\M)$. \\
    \If{$\alpha(S_k)\geq 1/\log n$ or $|S_k|>n/2$} {
        \Return{$S_1, \dots ,S_{k-1}$}
    }
    }
    \Return{$S_1, \dots ,S_k$}
\end{algorithm}

We first observe that since we invoked \cref{alg:RemoveSmallCircuits} to eliminate every circuit of size $\leq 50$ at the beginning, we always have $\alpha(S_i)\geq 50$. We provide the following characterization of how the $\alpha$-value of the sets changes:

\begin{claim}\label{clm:peelingSetsAlpha}
Let $\mathcal{M}$ be a matroid, and let $S_1, \dots S_k$ be a sequence of sets that are peeled off in accordance with \cref{alg:iterativePeel}. For any $i<j$ where $|S_i|\leq 2|S_j|$, we have
\[
\alpha(S_j) \geq \frac{\alpha(S_i)|S_j|}{|S_i|} + \frac{1}{5\sqrt{2}}\sqrt{\frac{\alpha(S_i)|S_j|}{|S_i|}}.
\]    
\end{claim}

\begin{proof}
Let $\ell = \frac{\alpha(S_i)}{|S_i|} n$. Suppose $U$ is a subset of the ground set $E$ of cardinality $\ell$, we define random variables $X_i = |U\cap S_i|, X_j =|U \cap S_j|$. Note that $X_i\sim \Hyp(n,|S_i|,\ell),X_j\sim \Hyp(n,|S_j|,\ell)$ and they are negatively correlated.

For the sake of contradiction, suppose
\[
\alpha(S_j) \leq \frac{\alpha(S_i)|S_j|}{|S_i|} +  \frac{1}{5\sqrt{2}} \sqrt{\frac{\alpha(S_i)|S_j|}{|S_i|}}.
\]
We aim to show that
\[
\Pr[X_i< \alpha(S_i) \land X_j\geq \alpha(S_j)] > 4\cdot 2^{-19}.
\]
Consider the $i$-th iteration of \cref{alg:iterativePeel}, the above implies that for a random permutation $\pi$ of $[n]$, there is a $>4\cdot 2^{-19}$ probability that there are less than $\alpha(S_i)$ elements from $S_i$ and more than $\alpha(S_j)$ elements from $S_j$ in the first $\ell$ elements of $\pi$. Conditioned on this, with at least $1/4$ probability, there is no circuit within $S_i$ but there is a circuit in $S_j$. This implies that with $>2^{-19}$ probability, the first circuit appears outside of $S_i$ when adding elements according to the order of a random permutation $\pi$. But on the other hand, since $S_i$ is a greedility-optimal set in the $i$-th iteration, by \cref{clm:boundedErrorEstimation}, we have $q(S) \geq 1-2^{-20}-1/n_0^2\geq 1-2^{-19}$. This is a contradiction.

To estimate $X_i\sim \Hyp(n,|S_i|,\ell)$, we define $Y_i\sim \Bin(\ell,|S_i|/n)$. It gives a good estimation as $\ell/n = \alpha(S_i)/|S_i|\leq 1/\log n$. Since $|S_i|\leq n/2$, we see that
\[
\E[Y_i] = \alpha(S_i),\quad \Var[Y_i] = \ell \frac{|S_i|}{n}\left(1-\frac{|S_i|}{n}\right) \geq  \frac{\alpha(S_i)}{2}\geq 25
\]
It follows from \cref{lem:binomial} that
\[
\Pr[Y_i<\alpha(S_i)] = \Pr[Y_i\leq \E[Y_i]-1] \geq \Pr\left[Y_i\leq \E[Y_i]-\frac{1}{5}\sqrt{\Var[Y_i]}\right]\geq \frac{1}{108}.
\]
Combined with \cref{lem:total-variation} and $\ell/n\leq 1/\log n$, we conclude that
\[
\Pr[X_i<\alpha(S_i)] \geq \frac{1}{108}-\frac{\ell-1}{n-1}\geq \frac{1}{200}.
\]

To estimate $X_j\sim \Hyp(n,|S_j|,\ell)$, we also define $Y_j\sim \Bin(\ell,|S_j|/n)$. Since $|S_j|\leq n/2, |S_i|\leq 2|S_j|$, we see that
\[
\E[Y_j]=\frac{\alpha(S_i)|S_j|}{|S_i|},\quad \Var[Y_j]=\ell \frac{|S_j|}{n}\left(1-\frac{|S_j|}{n}\right) \geq  \frac{\alpha(S_i)|S_j|}{2|S_i|}\geq \frac{\alpha(S_i)}{4} \geq 1.
\]
It follows from \cref{lem:binomial} that
\[
\Pr\left[Y_j\geq \E[Y_j]+\frac{1}{5}\sqrt{\Var[Y_j]}\right] \geq \frac{1}{108}
\]
Combined with \cref{lem:total-variation} and $\ell/n\leq 1/\log n$, we have
\[
\Pr\left[X_j\geq \E[Y_j]+\frac{1}{2}\sqrt{\Var[Y_j]}\right] \geq \frac{1}{108} - \frac{\ell-1}{n-1} \geq \frac{1}{200}.
\]
We conclude that
\[
\Pr[X_j\geq \alpha(S_j)]\geq \Pr\left[X_j\geq \frac{\alpha(S_i)|S_j|}{|S_i|}+\frac{1}{5\sqrt{2}}\sqrt{\frac{\alpha(S_i)|S_j|}{|S_i|}}\right] \geq \Pr\left[X_j\geq \E[Y_j]+\frac{1}{5}\sqrt{\Var[Y_j]}\right] \geq \frac{1}{200}.
\]
The penultimate inequality follows from
\[
\E[Y_j] = \frac{\alpha(S_i)|S_j|}{|S_j|},\quad \Var[Y_j]\geq\frac{\alpha(S_i)|S_j|}{2|S_i|}. 
\]

Since $X_i,X_j$ are negatively correlated, we have
\[
\Pr[X_i<\alpha(S_i)\land X_j\geq \alpha(S_j)] \geq \Pr[X_i<\alpha(S_i)] \cdot \Pr[X_j\geq \alpha(S_j)] \geq \frac{1}{200^2} > 4 \cdot 2^{-19}.
\]
as desired.
\end{proof}

Now, we will bound the growth of the $\alpha$ values as a function of the number of sets that are peeled off:

\begin{lemma}\label{lem:quadraticGrowth}
Let $\mathcal{M}$ be a matroid, and let $S_1, \dots S_k$ be a sequence of sets that are peeled off in accordance with \cref{alg:iterativePeel}. Now, let $\ell\in [\log n]$ be an integer, let $T = \{ i \in [k]: |S_i| \in [2^{\ell}, 2^{\ell+1} -1]\}$, let $\gamma = |T|$, and let $a_1, \dots a_{\gamma}$ denote the indices in $T$. Then it must be the case that 
\[
    \alpha(S_{a_{\gamma}}) = \Omega(\gamma^2), \quad \gamma = O\left(\sqrt{2^{\ell}}\right).
\]
\end{lemma}

\begin{proof}
As $|S_{a_i}|\leq 2|S_{a_{i+1}}|$, we have
\[
\alpha(S_{a_{i+1}}) \geq \frac{\alpha(S_{a_i})|S_{a_{i+1}}|}{|S_{a_i}|} + \frac{1}{5\sqrt{2}} \sqrt{\frac{\alpha(S_{a_i})|S_{a_{i+1}}|}{|S_{a_i}|}}.
\]
by \cref{clm:peelingSetsAlpha}. Now, we multiply both sides with $2^{\ell}/|S_{a_{i+1}}|$:

\[
\frac{\alpha(S_{a_{i+1}})}{|S_{a_{i+1}}|} \cdot 2^{\ell} \geq \frac{\alpha(S_{a_i})}{|S_{a_i}|} \cdot 2^{\ell} + \frac{1}{5\sqrt{2}} \sqrt{\frac{\alpha(S_{a_i})}{|S_{a_i}|} \cdot \frac{2^{\ell}}{|S_{a_{i+1}}|} \cdot 2^{\ell}} \geq \frac{\alpha(S_{a_i})}{|S_{a_i}|} \cdot 2^{\ell} + \frac{1}{10} \sqrt{\frac{\alpha(S_{a_i})}{|S_{a_i}|} \cdot 2^{\ell}}
\]
If we set $X_i = \frac{\alpha(S_{a_{i}})}{|S_{a_{i}}|}\cdot 2^{\ell}$. We get the relationship that 
\[
X_{i+1} = X_i + \frac{\sqrt{X_i}}{10}.
\]
As $X_1\geq 1$, the recurrence implies that $X_\gamma=\Omega(\gamma^2)$. Therefore, we conclude that
\[
\alpha(S_{a_{\gamma}}) = X_\gamma \cdot \frac{|S_{a_\gamma}|}{2^{\ell}} = \Omega(\gamma^2).
\]
As $\alpha(S_{a_\gamma}) \leq |S_{a_\gamma}|\leq 2^{\ell+1}$, we have $\gamma = O\left(\sqrt{2^{\ell}}\right)$.
\end{proof}

\begin{claim}
\label{clm:peel-round}
Let $\mathcal{M}$ be a matroid, and let $S_1, \dots S_k$ be a sequence of sets that are peeled off in accordance with \cref{alg:iterativePeel}. We have $k=O(n^{1/3})$.
\end{claim}
\begin{proof}
For every $\ell\in [\log n]$, we let $T_\ell= \{ i \in [k]: |S_i| \in [2^{\ell}, 2^{\ell+1} -1]\}$. It follows from \cref{lem:quadraticGrowth} that $|T_\ell| = O\left(\sqrt{2^{\ell}}\right)$. We can also trivially bound $|T_\ell| = O\left(\frac{n}{2^\ell}\right)$ as the total size of the sets is at most $n$. Therefore, we see that
\[
k=\sum_{\ell=1}^{\log n} |T_\ell| = \sum_{\ell=1}^{(2/3)\log n} |T_\ell| + \sum_{(2/3)\log n + 1}^{\log n} |T_\ell| = \sum_{\ell=1}^{(2/3)\log n} O(\sqrt{2^{\ell}}) + n^{1/3} = O(n^{1/3}).
\]
\end{proof}

Combining \cref{clm:peel-round} and \cref{lem:quadraticGrowth} gives the following theorem (which we state explicitly, as it may be of independent interest):

\begin{theorem}[Decomposition Theorem]
    Let $\mathcal{M}$ be a matroid, and let $S_1, \dots S_k$ be a sequence of sets that are peeled off in accordance with \cref{alg:iterativePeel}. Now, let $\ell\in [\log n]$ be an integer, let $T = \{ i \in [k]: |S_i| \in [2^{\ell}, 2^{\ell+1} -1]\}$, let $\gamma = |T|$, and let $a_1, \dots a_{\gamma}$ denote the indices in $T$. Then it must be the case that 
\begin{enumerate}
    \item $\alpha(S_{a_{\gamma}}) = \Omega(\gamma^2)$.
    \item $\gamma = O\left(\sqrt{2^{\ell}}\right)$.
    \item $k=O(n^{1/3})$.
\end{enumerate}
\end{theorem}

\section{Making Progress With Large $\alpha$'s}\label{sec:largeAlpha}

In this section, we will describe how we can make progress by \emph{contracting a large independent set} when we obtain a set $S$ with large $\alpha(S)$ during our decomposition. In the subsequent section, we will show that when the $\alpha(S)$ values are small, there are different ways of making progress. Collectively, these results will allow for a case-based analysis that allows us to beat $O(\sqrt{n})$ rounds no matter what is returned during the decomposition, which we present in \cref{sec:tradeoff}.

\begin{claim}\label{clm:parentMatroidalpha}
Let $\M$ be a matroid on $n$ elements, and let $S$ be a greedily-optimal set in $\M$. Then, for $\ell = \frac{\alpha(S)}{10|S|}n$, we have
\[
\Pr_\pi[\Ind(\{\pi(1),\dots,\pi(\ell)\})=1] \geq \frac{1}{4}.
\]
\end{claim}
\begin{proof}
Suppose $U$ is a random subset of $E$ of cardinality $\ell$. We define a random variable $X=|U\cap S|$. Note that $X\sim \Hyp(n,|S|,\ell)$, and $\E[X]=\alpha(S)/10$. By Markov's inequality,
\[
\Pr[X\geq \alpha(S)] \leq \frac{1}{10}.
\]
In the $i$-th iteration of \cref{alg:iterativePeel}, as $S$ is a greedily-optimal set, we have $q_{\M}(S)\geq 1-2^{-20}-1/n_0^2$ by \cref{clm:boundedErrorEstimation}. In particular, if there is no circuit in $S \cap U$, this bounds the probability of a circuit appearing outside of $S$ by $2^{-20}+1/n_0^2$. Using this, we obtain:

\begin{align*}
\Pr_\pi[\Ind(\{\pi(1),\dots,\pi(\ell)\})= 0]  &\leq \frac{1}{2^{20}} + \frac{1}{n_0^2}+\Pr_U[\Ind(U\cap S)=0]\\
& \leq \frac{1}{2^{20}}+ \frac{1}{n_0^2}+\left(\Pr_{T\in {S\choose X}}[\Ind(T)=0\mid X< \alpha(S)] + \Pr[X\geq \alpha(S)]\right)\\
& \leq \frac{1}{2^{20}}+ \frac{1}{n_0^2}+\frac{1}{2}+\frac{1}{10} \leq \frac{3}{4}.    
\end{align*}
The first line of the proof follows by seeing that if there is a dependence in $\{\pi(1),\dots,\pi(\ell)\}$, then either (1) this is a circuit in $U \cap S$, or (2) there is no circuit in $U \cap S$, but there is a circuit outside $U \cap S$. Our bound on $q_{\M}(S)$ states that the probability of this second case is bounded by $\frac{1}{2^{20}} + \frac{1}{n_0^2}$. This concludes the proof. 
\end{proof}

Given the above claim, it follows that we can recover an independent set of size $\Omega\left(\frac{\alpha(S)}{|S|}n\right)$ with probability at least $1-2^{-n_0}$ by sampling $\text{poly}(n_0)$ many random permutations.

\section{Making Progress With Small $\alpha$'s}\label{sec:redundant}

In this section, we will describe how we can make progress on \emph{deleting redundant elements} when we obtain a set $S$ with a small $\alpha(S)$ value. The reader may see \cref{sec:matroidTheoryPrelims} for the definitions of some of the quantities we make use of in this section. 

\subsection{Rank Deficiency in Matroids}

Our first result is a general structural result which governs the so-called \emph{rank deficiency} of matroids.

\begin{definition}
    For a matroid $\M$ on $n$ elements of rank $r$ we say that the rank-deficiency is $n - r$.
\end{definition}

\subsubsection{Matroid Decomposition for Quotient Counting Bounds}
Now, recall the work of \cite{Qua23} showed the following key theorem:
\begin{theorem}\label{thm:quotientBound}[Lemma 3.2 in \cite{Qua23}]
Let $\M=(E,\I)$ be a matroid of rank $r$ with $|E|=n$, and let 
\[
    \kappa(\M) = \min_{S \subseteq E} \frac{n - |S|}{r - \mathrm{rank}(S)}.
\]
Then, for any $\eta \in \Z^+$, there are $\leq n^{\eta+1} r^{\eta}$ quotients of $\M$ of size $\leq \eta \cdot \kappa(\M)$.
\end{theorem}

\begin{remark}
    The expression 
    \[
    \kappa(\M) = \min_{S \subseteq E} \frac{n - |S|}{r - \mathrm{rank}(S)}
    \]
    is always minimized for $S = \mathrm{span}(S)$, i.e. $S$ is a flat. Otherwise, WLOG we can set $S = \mathrm{span}(S)$, and the denominator is unchanged while the numerator decreases.
\end{remark}

Now, we have the following key claim:

\begin{claim}\label{clm:decomposeIncreaseKappa}
Let $\M=(E,\I)$ be a matroid of rank $r$ with $|E|=n$ and let $d$ be a given parameter. Then, there exists a set $T$ of size $|T|\leq r\cdot d$ such that the new matroid $\M \setminus T$ is either empty or satisfies $\kappa(\M \setminus T) \geq d$.
\end{claim}

\begin{proof}
Consider the following iterative process: We start with $T=\emptyset$ and $\M_1=\M$. In the $i$-th iteration:
\begin{itemize}
    \item If $\M_i=\emptyset$ or $\kappa(\M_i)\geq d$, we are done.
    \item Otherwise, let $\M_i=(E_i,\I_i)$. It implies there exist a flat $S\subseteq E_i$ such that
    \[
    \frac{|E_i|-|S|}{\text{rank}(\M_i)-\text{rank}(S)} <d.
    \]
    We set $\M_{i+1} \gets \M_{i}|_S,T\gets T\cup (E_i\setminus S)$. Note that $\rank(\M_{i+1})=\rank(S)$ and the size of $T$ has increased by $|E_i|-|S|<d\cdot (\text{rank}(\M_i)-\text{rank}(S))=d \cdot (\text{rank}(\M_i)-\text{rank}(\M_{i+1}))$.
\end{itemize}
Suppose the process terminates after $t$ iterations. We finish the proof by observing that
\[
|T|< \sum_{i=1}^t d(\rank(\M_i)-\rank(\M_{i+1})) = d(\rank(\M_1)-\rank(\M_t)) \leq d\cdot \rank(\M_1) = d\cdot r.
\]
\end{proof}

\begin{theorem}\label{thm:quotientBoundDecomposition}
Let $\M=(E,\I)$ be a matroid on $n$ elements of rank $r$, and let $d \in \Z^+$ be a parameter of our choosing. Then, there is a set of elements $T \subseteq E$ of size $|T| \leq r \cdot d$ such that the matroid $\M'=\M\setminus T$ is either empty, or for any $\eta \in \Z^+$, $\M'$ has at most $n^{2 \eta+1}$ quotients of size $\leq \eta d$.
\end{theorem}

\begin{proof}
This follows by letting $T$ be the set of elements recovered by \cref{clm:decomposeIncreaseKappa}, and then invoking \cref{thm:quotientBound} on the matroid $\M \setminus T$ (if non-empty), where $\kappa(\M \setminus T) \geq d$.
\end{proof}

\subsubsection{Proof of Rank Deficiency}

With all of this groundwork established, we are now ready to prove our theorem for general matroids:

\begin{theorem}\label{thm:rankDeficiency}
Let $\M=(E,\I)$ be a matroid on $n$ elements. For a set $S\subseteq E$ and matroid $\M'=\M|_S$ such that
\begin{enumerate}
    \item $\alpha(S) \leq \frac{|S|}{100\log n}$.
    \item For every element $x \in S$, $p_{\M'}(x) \geq \frac{1}{n^{10}}$.
\end{enumerate}
    Then, $|S| - \mathrm{rank}(S) = \Omega(|S| / \log n)$, where $\mathrm{rank}(S) = \mathrm{rank}(\M')$.
\end{theorem}

\begin{proof}
We invoke \cref{thm:quotientBound} on the dual matroid $(\M')^*$, with parameter $d = 100 \log n$. The result tells us that there is a set $T \subseteq S$ of size $|T|\leq (|S|-\rank(S)) 100 \log n$ such that the resulting matroid $(\M')^*\setminus T$ is either empty or satisfies a quotient counting bound: specifically, that the number of quotients of size $\leq \eta d$ is at most $|S|^{2 \eta+1}$. We show that it must be the former case by showing a contradiction assuming $(\M')^*\setminus T$ is non-empty.

In particular, let us now consider sampling this matroid $(\M')^*\setminus T$ at rate $1/2$. Our goal is to bound the probability that there exists \emph{any quotient} which survives (here, we take surviving to mean that \emph{every} element in the quotient is selected during the sampling) this sampling procedure. We see that this is bounded by
\begin{align*}
\Pr[\exists c \in Q((\M')^*\setminus T): c \text{ survives sampling}] & \leq \sum_{c \in Q((\M')^*\setminus T) } \left ( \frac{1}{2} \right )^{|c|}\\
& \leq \sum_{\eta \in \Z^+} n^{2\eta+1} \cdot \left ( \frac{1}{2}\right )^{\eta d}\\
& \leq n\cdot\sum_{\eta \in \Z^+} \frac{1}{n^{98 \eta}} \leq \frac{2}{n^{97}}.
\end{align*}

It follows from \cref{fact:dualDeletion} that $((\M')^*\setminus T)^* = \M'/T$. By \cref{fact:circuitHyperplane}, we know that any circuit $C$ of $\M'/T$ is the complement to a hyperplane of $(\M'/T)^* = (\M')^*\setminus T$. In particular, because hyperplanes are flats, we know that any circuit of $\M'/T$ is a quotient of $(\M')^*\setminus T$. Importantly, this means that when we sample the matroid $\M'/T$ at rate $1/2$, there is a $\leq 2 / n^{97}$ chance of any circuit surviving. 

Now, consider the elements $S\setminus T$. Recall that in the original matroid, we know that $\forall x \in S$, $p_{\M'}(x) \geq \frac{1}{n^{10}}$. In particular, because $\alpha(S) \leq \frac{|S|}{100\log n}$, if we sample $\M'$ at rate $1/2$, we expect $\geq 50\log n \cdot \alpha(S)$ elements to survive the sampling. It follows from a simple Chernoff bound that with probability at least $1-n^{-5}$, we have  $\geq 25\log n\cdot \alpha(S)$ elements that are selected during sampling. Conditioned on this, by \cref{clm:alpha-dependent}, we see that a circuit will survive with probability at least $1-n^{-25}$. 

The final key observation is the following: if $T \neq S$, there must exist elements $x \in S \setminus T$. Originally, these elements satisfied $p_{\M'}(x)\geq \frac{1}{n^{10}}$. Thus, if we sample the original matroid $\M'$ at rate $1/2$, we would have that the probability that there is a circuit in the resulting sample which includes $x$ is at least
\[
\left(1-\frac{1}{n^{5}}\right)\left(1-\frac{1}{n^{25}}\right) \frac{1}{n^{10}} \geq \frac{1}{n^{11}}.
\]
However, we are now no longer sampling the matroid $\M'$, but rather sampling the matroid $\M' / T$. For analysis, we consider correlating the two sampling procedures; i.e., we let $P$ denote the sample of elements received in $\M'$, and we let $\widetilde{P}$ denote the same set of sampled elements in $\M' / T$, so $\widetilde{P} = P \setminus T$. The point now is that if there is a circuit $C \subseteq P$ such that $C \setminus T\neq \emptyset$, \emph{there must also be a circuit $\widetilde{C} \subseteq C\setminus T \subseteq \widetilde{P}$}. To see this, we note that $\rank_{\M'}(C)=|C|-1$ as $C$ is a circuit. Likewise, $\rank_{\M'}(C\cap T)=|C\cap T|$ as $C\cap T \subsetneq T$ and thus it must be independent (otherwise it contracts that $C$ is a minimal dependent set). By \cref{fact:dualDeletion}, we have
\begin{align*}
\rank_{\M'/T}(C\setminus T) & = \rank_{\M'}((C\setminus T)\cup T) - \rank_{\M'}(T)\\
& = \rank_{\M'}(C\cup T) - \rank_{\M'}(T)\\
& \leq \rank_{\M'}(C)+\rank_{\M'}(T)-\rank_{\M'}(C\cap T)-\rank_{\M'}(T)\\
& \leq |C|-1 - |C\cap T|\\
&  \leq |C\setminus T|-1.
\end{align*}
The first inequality follows from the submodularity of the rank function of a matroid. Therefore, we conclude that $C\setminus T$ is dependent in $\M'/T$ and there is a circuit inside it.

But this leads to a contradiction that
\begin{align*}
    \frac{2}{n^{97}} & \geq \Pr[\exists \text{ circuit in }\M'/T \text{ when sampling with rate }1/2]\\
    & \geq \Pr[\exists \text{ circuit } C \text{ in }\M '\text{ when sampling with rate }1/2 \text{ s.t. } C\setminus T\neq \emptyset] \geq \frac{1}{n^{11}}
\end{align*}

In particular, this means that the matroid $(\M')^*\setminus T$ must be empty. Therefore $|S|=|(\M')^*| = |T| \leq (|S|-\rank(S))100\log n$, and so $|S|-\rank(S) = \Omega(|S| / \log n)$.
\end{proof}

\subsection{Efficient Redundant Element Recovery}

In this section, we present an algorithm which can efficiently recover redundant elements, provided the $\alpha$ values are significantly smaller than the set size. 

\begin{algorithm}
    \caption{RecoverRedundantElements$(\M,S)$}\label{alg:recover-reduant-elements}
    Let $\widehat\alpha(S)$ be the estimation of $\alpha(S)$ given by \cref{clm:alpha-estimation}\\
    $t\gets 20\log n\cdot \widehat\alpha(S), \ell \gets \frac{|S|}{4t}$\\
    \For{$i \in [\ell]$ in parallel} {
        Draw a random permutation (bijection) $\pi:[|S|]\to S$\\
        $A_i \gets \{\pi(1),\dots, \pi(t)\},B_i\gets \emptyset$\\
        \For{$j \in [t]$ in parallel}{
            \For{$x\in E\setminus A_i$ in parallel} {
                Query $\Ind(\{\pi(1),\dots,\pi(j)\})$ and $\Ind(\{\pi(1),\dots,\pi(j)\}\cup \{x\})$\\
                \If{$\Ind(\{\pi(1),\dots,\pi(j)\})=1\land\Ind(\{\pi(1),\dots,\pi(j)\}\cup \{x\}) =0$} {
                    $B_i\gets B_i\cup \{x\}$
                }
            }
        }
    }
    \Return{$\bigcup_{i\in[\ell]} B_i \setminus (\bigcup_{i\in[\ell]} A_i)$}
\end{algorithm}

We establish the following lemma:

\begin{lemma}\label{lem:efficientRedundant}
Let $\M$ be a matroid, and $S$ be a greedily-optimal set and $\alpha(S)\leq \frac{|S|}{100\log n}$. Then, \cref{alg:recover-reduant-elements} recovers $\widetilde\Omega \left(\min\left\{|S|,\frac{|S|^2}{\alpha(S)^2}\right\}\right)$ redundant elements with probability $1 - 1 / 2^n$.
\end{lemma}

\begin{proof}
First, we note that
\[
\left(\bigcup_{i\in[\ell]} B_i \setminus \left(\bigcup_{i\in[\ell]} A_i\right)\right) \subseteq \text{span}\left(\bigcup_{i\in[\ell]} A_i\right)
\]
Thus, the recovered set is indeed redundant. We now focus on bounding the size of this set in the following.

As $S$ is a greedily-optimal set in $\M$, by \cref{clm:highProbCircuit}, we have for any $x\in S$, $p_{\M|_S}(x)\geq \frac{1}{2^{21}|S|\log n}$. For every $x\in S$, we say that $p_{x,r}$ is the probability over a random order of picking elements such that $x$ is the $r$-th element added, that $x$ participates in the first circuit that appears.

In particular, we can establish some simple inequalities:
\begin{enumerate}
    \item $p_{x,r} \geq p_{x,r + 1}$.
    \item $p_{\M|_S} (x) = \frac{1}{|S|} \sum_{r = 1}^{|S|} p_{x,r}$.
    \item $\sum_{r = t+1}^{|S|} p_{x,r} \leq 1 / n^{10}$.
\end{enumerate}
The third inequality follows from \cref{clm:alpha-dependent}: a random subset of $S$ of cardinality $t=20\widehat\alpha(S)\log n\geq 10\alpha(S)\log n$ is dependent with probability at least $1-1/n^{10}$.

With this, we can see that for any $x\in S$,
\[
    \frac{1}{2^{21}|S|\log n}\leq p_x =\frac{1}{|S|} \sum_{r = 1}^{|S|} p_{x,r} = \frac{1}{|S|}\left( \sum_{r = 1}^{t} p_{x,r} + \frac{1}{n}\right) \leq \frac{t}{|S|} \cdot p_{x,1} + \frac{1}{n^{10}}.
\]
In particular, this implies that 
\[
    p_{i,1} \geq \frac{|S|}{t} \cdot \left(\frac{1}{2^{21}|S|\log n} - \frac{1}{n^{10}}\right) \geq\frac{1}{2^{22}t\log n}
\]

Now, let us revisit the above algorithm. Our first step will be to understand the probability that an element $x$ appears in one of the sets $A_1, \dots A_{\ell}$. For this, observe that each set $A_i$ is of size $t$. Thus, 
\[
    \Pr\left[x \notin \bigcup_{i\in[\ell]} A_i \right] = \Pr[x \notin A_1]^{\ell} = \left ( 1 - \frac{t}{|S|}\right )^{\ell} =\left ( 1 - \frac{t}{|S|}\right )^{\frac{|S|}{4t}} \geq  e^{-1/2} \geq 1/2.
\]
The first inequality is because $\frac{t}{|S|} = \frac{20 \log n \cdot \widehat \alpha(S)}{|S|}\leq \frac{40\log n\cdot \alpha(S)}{|S|} \leq \frac{1}{2}$ and for every $0\leq x\leq \frac{1}{2}, 1-x\geq e^{-2x}$.

Now, let us introduce the value $q_i$ such that 
\[
    q_x = \Pr\left[x \notin \bigcup_{i\in[\ell]} A_i \wedge x \in \bigcup_{i\in[\ell]} B_i\right] = \Pr\left[x \notin \bigcup_{i\in[\ell]} A_i\right] \cdot \Pr\left[x \in \bigcup_{i\in[\ell]} B_i\;\middle|\; x \notin \bigcup_{i\in[\ell]} A_i\right].
\]
Note that the samples $A_1, \dots A_{\ell}$ are all done independently of one another. Hence,
\[
\Pr\left[x \in \bigcup_{i\in[\ell]} B_i\;\middle|\; x \notin \bigcup_{i\in[\ell]} A_i\right] = 1 - \Pr\left[x \notin \bigcup_{i\in[\ell]} B_i\;\middle|\; x \notin \bigcup_{i\in[\ell]} A_i\right] = 1-\Pr[x\not\in B_1\mid x\notin A_1]^{\ell}.
\]
Now, let us understand $\Pr[x\in B_1 | x\notin A_1]$. This is exactly the probability of $x$ appearing in the first circuit when we randomly add the set $A_1$ of elements to $x$. Since $A_1$ is disjoint from $x$, this is exactly $p_{i,1}$. Hence, we obtain that
\begin{align*}
1 - \Pr[x \notin B_1 | x \notin A_1]^{\ell} & = 1 - (1 - p_{x,1})^{\ell}\\
& \geq 1 - \left ( 1 - \frac{1}{2^{22}t\log n}\right)^{\frac{|S|}{4t}} \\
& \geq 1-\exp\left(\frac{|S|}{2^{24}t^2\log n}\right)\\
& \geq \left\{\frac{1}{2},\frac{|S|}{2^{25}t^2\log  n}\right\}
\end{align*}
The last inequality follows from the fact that $1-e^{-x}\geq \min\{1/2,x/2\}$ when $x\geq 0$.

To conclude, we obtain that 
\[
q_x \geq \frac{1}{2} \cdot (1 - \Pr[x \notin B_1 | x \notin A_1]^{\ell}) \geq \min\left \{\frac{1}{4}, \frac{|S|}{2^{26}t^2\log n} \right \}
\]
Finally then, we see that 
\begin{align*}
\E\left[\left|\bigcup_{i\in[\ell]} B_i \setminus \left(\bigcup_{i\in[\ell]} A_i\right)\right|\right] & = \sum_{x\in S} q_x\\
& \geq \sum_{x\in S}\min\left \{\frac{1}{4}, \frac{|S|}{2^{26}t^2\log n} \right \}\\
& \geq \min\left \{\frac{|S|}{4}, \frac{|S|^2}{2^{26}t^2\log n} \right \}\\
& = \widetilde\Omega \left(\min\left\{|S|,\frac{|S|^2}{\widehat\alpha(S)^2}\right\}\right)\\
& = \widetilde\Omega \left(\min\left\{|S|,\frac{|S|^2}{\alpha(S)^2}\right\}\right).
\end{align*}
Repeating the above $\text{poly}(n_0)$ times achieves at least this expectation with probability at least $1 - 2^{-n_0}$ by a Hoeffding's inequality.

\end{proof}

\section{Guaranteeing Progress through Decomposition}\label{sec:tradeoff}

We now establish several warm-up claims about when it is easy to make progress during our decomposition. First, to establish uniform notation, we consider the following process:

\begin{algorithm}
    \caption{EarlyStopDecomposition$(\M)$}\label{alg:earlyStop}
    $\M\gets \text{RemoveSmallCircuits}(\M)$\\
    $k \gets 0$ \\
    \While{$|E| \geq n/2$}{
        $k\gets k+1$ \\
        $S_k \gets \mathrm{Peel}(\M)$ \\
        $\M \gets \M \setminus S_k$. \\
        Let $\widehat\alpha(S_k)$ be the estimation of $\alpha(S_k)$ given by \cref{clm:alpha-estimation}\\
        \If{$\widehat \alpha(S_k)=\Omega(|S_k|/\log n)$} {
            \Return{$k-1, S_1, \dots S_{k-1}, \M$} \label{line:alpha-return}
        }
    }
    \Return{$k, S_1, \dots S_k, \M$}
\end{algorithm}

\subsection{Subroutines for Making Progress Towards a Basis}
We define
\[
i^* = \arg\max_{i\in[k]} \frac{\alpha(S_i)}{|S_i|}.
\]
For $\ell\in[\log n]$, let $J_\ell = \{i\in[k]\mid |S_i|\in[2^{\ell},2^{\ell+1}-1]\}$. We see that at least one $J_{\ell^*}$ satisfies $|J_{\ell^*}|\geq k/\log n$. We denote the sets $S_i$ for $i\in J_{\ell^*}$ by $T_1,\dots,T_{\gamma}$, and let $\tau=2^{\ell^*},\beta=\tau\cdot\alpha(S_{i^*})/|S_{i^*}|$.

\begin{claim}\label{clm:propertiesTauBeta}
We have the following properties:
\begin{enumerate}
    \item $\gamma = \widetilde\Omega(k)$.
    \item For any $i\in [\gamma]$, $\tau\leq |T_i|\leq 2\tau$.
    \item For any $i\in [\gamma]$, $\alpha(T_i)\leq \frac{|T_i|}{100\log n}$.
    \item For any $i\in [k]$, $\frac{\alpha(S_i)}{|S_i|} \leq \frac{\beta}{\tau}$. For any $i\in [\gamma]$, $\alpha(T_i)=O(\beta)$.
    \item $\beta = \Omega(\gamma^2), \tau = \Omega(\beta)$.
\end{enumerate}
\end{claim}

\begin{proof}
    Item 1 follows from our notational choices, as $J_{\ell^*}$ satisfies $\gamma = |J_{\ell^*}|\geq k/\log n$. Item $2$ again follows by our definition of $\tau$. Item 3 is because if $\alpha(T_i)>\frac{|T_i|}{100 \log n}$, the algorithm would have returned on \cref{line:alpha-return} as $\widehat \alpha(T_i) = \Theta(\alpha(T_i))=\Omega(|T_i|/\log n)$. Item 4 follows from the maximality of $\alpha(S_{i^*})/|S_{i^*}|$ and $\alpha(T_i)\leq \frac{\beta}{\tau} |T_i| = O(\beta)$. For item $5$, it follows from \cref{lem:quadraticGrowth} that $\alpha(T_{\gamma})=\Omega(\gamma^2)$. Given $\alpha(T_\gamma)=O(\beta)$ by item 3, we see that $\beta=\Omega(\gamma^2)$. Additionally, we have $\tau=\Omega(\beta)$ as $\frac{\beta}{\tau} = \frac{\alpha(S_k)}{S_k} < 1$.
\end{proof}

\begin{claim} \label{clm:return-recover}
If the algorithm returns on \cref{line:alpha-return}, then there is an efficient procedure for recovering an independent set of size $\widetilde \Omega(n)$ with probability $1-2^{\Omega(n_0)}$ by investing a single additional round with $\mathrm{poly}(n_0)$ queries.
\end{claim}
\begin{proof}
Let $\M$ be the matroid before we peel off $S_{k}$. Since $S_{k}$ is a greedily-optimal set in $\M$ and $\alpha(S_k) = \Theta(\widehat \alpha(S_k)) = \Omega(|S_k|/\log n)$, by \cref{clm:parentMatroidalpha}, for $\ell = \frac{n\alpha(S_k)}{10|S_k|} = \widetilde \Omega(n)$,
\[
\Pr_{\pi} [\Ind(\{\pi(1),\dots,\pi(\ell)\}) =1] \geq \frac{1}{4}.
\]
Thus, by sampling $\text{poly}(n_0)$ many random permutations, we can find an independent set of size $\ell$ with high probability in 1 additional round.
\end{proof}

By the above claim, we see that if the algorithm returns on \cref{line:alpha-return}, we are in a good situation since we have invested $O(k)=O(n^{1/3})$ rounds (\cref{clm:peel-round}) and can recover an independent set of size $\widetilde \Omega(n)$. Therefore, we assume that the algorithm \emph{does not} return on \cref{line:alpha-return} in the following. We present 4 different ways of making progress.

\begin{lemma}\label{lem:subroutine1}
There is an efficient procedure for recovering an independent set of size $\Omega\left(n\beta/\tau\right)$ with probability $1-2^{\Omega(n_0)}$ by investing a single additional round with $\mathrm{poly}(n_0)$ queries.
\end{lemma}

\begin{proof}
Let $\M$ be the matroid right before we peel off $S_{i^*}$. Since $S_{i^*}$ is a greedily-optimal set in $\M$ and $\alpha(S_{i^*})/|S_{i^*}| = \beta/\tau$, by \cref{clm:parentMatroidalpha}, for $\ell = \frac{n\beta}{10\tau}$
\[
\Pr_{\pi} [\Ind(\{\pi(1),\dots,\pi(\ell)\}) =1] \geq \frac{1}{4}.
\]
Thus, by sampling $\text{poly}(n_0)$ many random permutations, we can find an independent set of size $\ell$ with high probability in 1 additional round.
\end{proof}

\begin{lemma}\label{lem:subroutine2}
There is an efficient procedure for deleting $\widetilde\Omega\left(\gamma \tau\right)$ redundant elements by investing $ O\left(\tau^{1/2}\right)$ additional rounds and making only $\mathrm{poly}(n_0)$ queries.
\end{lemma}
\begin{proof}
For any $i\in [\gamma]$, we have $\alpha(T_i)\leq \frac{|T_i|}{100\log n}$. Thus, by \cref{thm:rankDeficiency}, $|T_i|-\rank(T_i) = \widetilde \Omega(|T_i|)$ for every $i\in [\gamma]$. Therefore, we can use the $O(n^{1/2})$ round algorithm of \cite{kuw85} to find an independent set $I_i$ for each $T_i$ in parallel in $O(|T_i|^{1/2})=O(\tau^{1/2})$ additional rounds. In total, we can remove
\[
\sum_{i=1}^{\gamma} |T_i|-|I_i| = \sum_{i=1}^\gamma \widetilde\Omega(|T_i|) = \widetilde\Omega(\gamma\tau)
\]
redundant elements.
\end{proof}

\begin{lemma}\label{lem:subroutine3}
There is an efficient procedure for deleting $\widetilde\Omega\left(\gamma \cdot \min\left(\tau,\frac{\tau^2}{\beta^2}\right)\right)$ redundant elements with probability $1 - 2^{-\Omega(n_0)}$ by investing a single additional round with $\mathrm{poly}(n_0)$ queries.
\end{lemma}

\begin{proof}
For every $i\in[\gamma]$, we have $\alpha(T_i)\leq \frac{|T_i|}{100 \log n}$. Therefore, we can invoke \cref{lem:efficientRedundant} (in parallel) across all $T_i$'s for $i\in[\gamma]$. This lemma guarantees that for each $T_i$, we recover
\[
    \widetilde\Omega\left (\min\left\{|T_i|, \frac{|T_i|^2}{\alpha(T_i)^2}\right\} \right ) \geq \widetilde\Omega\left (\min\left\{\tau, \frac{\tau^2}{\beta^2}\right\} \right ).
\]
In total, we recover
\[
\sum_{i\in [\gamma]} \widetilde\Omega\left (\min\left\{\tau, \frac{\tau^2}{\beta^2}\right\} \right ) =\widetilde\Omega\left (\gamma\cdot \min\left\{\tau, \frac{\tau^2}{\beta^2}\right\} \right ). 
\]
redundant elements.
\end{proof}

\begin{lemma}\label{lem:subroutine4}
There is an efficient procedure for deleting $\widetilde\Omega\left(\min\left(n,\frac{\tau^2}{\beta^2}\right)\right)$ redundant elements with probability $1 - 2^{-\Omega(n_0)}$ by investing a single additional round with $\mathrm{poly}(n_0)$ queries.
\end{lemma}

\begin{proof}
We apply \cref{lem:efficientRedundant} (in parallel) across every $S_i$ for $i \in [k]$.  As in the previous lemma's proof, \cref{lem:efficientRedundant} guarantees that we can recover 
\[
    \tilde\Omega\left (\min\left\{|S_i|, \frac{|S_i|^2}{\alpha(S_i)^2}\right\} \right ) \geq \tilde\Omega\left (\min\left\{|S_i|, \frac{\tau^2}{\beta^2}\right\} \right )
\]
redundant elements from each $S_i$. In total, we recover 
    \[
    \sum_{i \in [k]} \tilde\Omega\left (\min\left(|S_i|, \frac{\tau^2}{\beta^2}\right) \right ) = \tilde\Omega\left(\min\left(n, \frac{\tau^2}{\beta^2}\right)\right)
    \]
    redundant elements then. Note that this follows because in the sum, either in every term $|S_i| \leq \frac{\tau^2}{\beta^2}$, in which case the sum behaves like $\sum_i |S_i| = \Omega(n)$, or for at least one term $|S_i| > \frac{\tau^2}{\beta^2}$, and so the sum contains at least one contribution of $\frac{\tau^2}{\beta^2}$.
\end{proof}

\subsection{Piecing the Subroutines Together}

To summarize, we have 4 ways of making progress:
\begin{center}
\begin{tabular}{|c||c|c|}
\hline
& Progress & Round Complexity\\
\hline
\cref{lem:subroutine1} & $\widetilde\Omega\left(n\beta/\tau\right)$ & $\widetilde O(\gamma)$ \\ 
\hline
\cref{lem:subroutine2} & $\widetilde\Omega\left(\gamma\tau\right)$ & $\widetilde O(\gamma+\tau^{1/2})=\widetilde O(\tau^{1/2})$ \\ 
\hline
\cref{lem:subroutine3} & $\widetilde\Omega\left(\gamma \cdot \min\left(\tau,\frac{\tau^2}{\beta^2}\right)\right)$ & $\widetilde O(\gamma)$ \\ 
\hline
\cref{lem:subroutine4} & $\widetilde\Omega\left(\min\left(n,\frac{\tau^2}{\beta^2}\right)\right)$ & $\widetilde O(\gamma)$ \\ 
\hline
\end{tabular}
\end{center}

\begin{remark}\label{rmk:gammabeta}
    Note that we know $\gamma= O(\beta^{1/2}) = O(\tau^{1/2})$ because of \cref{clm:propertiesTauBeta}.
\end{remark}

Algorithmically, we will always choose the one that maximizes the average progress per round, which is given by
\[
\max\left\{\frac{n\beta}{\tau\gamma},\gamma\tau^{1/2},\min\left\{\tau,\frac{\tau^2}{\beta^2}\right\},\min\left\{\frac{n}{\gamma},\frac{\tau^2}{\beta^2\gamma}\right\}\right\}.
\]
Notationally, we will let $\mathbf{Progress}_1(\beta, \tau, \gamma), \mathbf{Rounds}_1(\beta, \tau, \gamma)$ denote the progress and round complexity guaranteed by \cref{lem:subroutine1}, $\mathbf{Progress}_2(\beta, \tau, \gamma), \mathbf{Rounds}_2(\beta, \tau, \gamma)$ that of \cref{lem:subroutine2}, \\$\mathbf{Progress}_3(\beta, \tau, \gamma), \mathbf{Rounds}_3(\beta, \tau, \gamma)$ that of \cref{lem:subroutine3}, and $\mathbf{Progress}_4(\beta, \tau, \gamma), \mathbf{Rounds}_4(\beta, \tau, \gamma)$ that of \cref{lem:subroutine4}.

Now, we have the following lemma:
\begin{lemma}\label{lem:caseBreakdown}
    Let $\beta,\tau,\gamma$ be the parameters resulting from running \cref{alg:earlyStop}. Then, for any possible $\beta,\tau,\gamma$, there is a choice of sub-routine $q \in [4]$ such that 
    \[
    \frac{\mathbf{Progress}_q( \beta, \tau, \gamma)}{\mathbf{Rounds}_q(\beta, \tau, \gamma)} = \widetilde{\Omega}(n^{8/15}).
    \]
\end{lemma}

\begin{proof}
We use a case analysis based on the values of $\tau, \beta$.
\begin{enumerate}
    \item When $\beta\geq n^{2/5}$:
        \begin{itemize}
            \item If $\tau\leq \beta^2/n^{2/15}$, then we immediately have that $\frac{\beta}{\sqrt{\tau}} \geq n^{1/15}$. Thus, we can write that
            \[
            n^{8/15} \leq n^{7/15} \cdot \frac{\beta}{\sqrt{\tau}},
            \]
            which implies that
            \[
            \frac{n^{8/15}}{\tau^{1/2}} \leq \frac{n^{7/15}\beta}{\tau}.
            \]
            Now again, we do a case analysis based on the value of $\gamma$: if $\gamma \geq\frac{n^{8/15}}{\tau^{1/2}} $, then 
            \[
            \frac{\mathbf{Progress}_2(\beta, \tau, \gamma)}{\mathbf{Rounds}_2(\beta, \tau, \gamma)} = \widetilde\Omega(\gamma\tau^{1/2}) = \widetilde\Omega( n^{8/15}).
            \]
            Otherwise, if $\gamma \leq \frac{n^{7/15}\beta}{\tau}$:
            \[
             \frac{\mathbf{Progress}_1(\beta, \tau, \gamma)}{\mathbf{Rounds}_1(\beta, \tau, \gamma)} = \widetilde\Omega\left(\frac{n\beta}{\tau\gamma}\right)\geq n^{8/15}.
            \]
            \item Now, we consider when $\tau\geq\beta^2/n^{2/15}\geq n^{2/3}\geq n^{8/15}$. Immediately, this implies that
            \[
            \frac{\tau^2}{\beta^2}\geq \frac{\beta^2}{n^{4/15}}\geq n^{8/15}.
            \]
            Thus, we have
            \[
             \frac{\textbf{Progress}_3( \beta, \tau, \gamma)}{\textbf{Rounds}_3(\beta, \tau, \gamma)} = \widetilde\Omega\left ( \min\left\{\tau,\frac{\tau^2}{\beta^2}\right\}\right) = \widetilde\Omega(n^{8/15}).
            \]
        \end{itemize}
    \item When $n^{2/15}\leq \beta\leq n^{2/5}$:
        Because $n^{1/15} \leq \sqrt{\beta} \leq n^{1/5}$, this means that
        \[
            n^{4/15}\beta \leq n^{7/15}\beta^{1/2} 
        \] and 
        \[
        n^{8/15}\leq n^{7/15}\beta^{1/2}.
        \]
        Now, again we do a case analysis. If $\frac{n\beta^{1/2}}{\tau}\geq n^{8/15}$, then 
        \[
        \tau\leq n^{7/15}\beta^{1/2}, 
        \]
        which means that 
        \[
        \frac{\mathbf{Progress}_1(\beta, \tau, \gamma)}{\mathbf{Rounds}_1(\beta, \tau, \gamma)} = \widetilde\Omega\left(\frac{n\beta}{\tau\gamma}\right) = \widetilde \Omega \left ( \frac{n \beta^{1/2}}{\tau} \right) \geq \widetilde \Omega \left ( n^{8/15}\right),
        \]
        where in the second equality we have used that $\gamma = O(\sqrt{\beta})$ as per \cref{rmk:gammabeta}. Now, in the second case, we consider what happens if $\frac{n\beta^{1/2}}{\tau} \leq n^{8/15} \leq n^{7/15} \beta^{1/2}$. Then, we obtain that $\frac{n}{\tau} \leq n^{7/15}$, which means $\tau \geq n^{8/15}$. Likewise, because $\beta \leq n^{2/5}$, this means that 
        \[
        \frac{\tau^2}{\beta^2}\geq n^{8/15}.
        \]
        Together, this implies that
        \[
        \frac{\mathbf{Progress}_3(\beta, \tau, \gamma)}{\mathbf{Rounds}_3(\beta, \tau, \gamma)} = \widetilde \Omega \left ( \min\left\{\tau,\frac{\tau^2}{\beta^2}\right\} \right) = \widetilde \Omega( n^{8/15}). 
        \]
    \item Finally, we consider what happens when $\beta\leq n^{2/15}$. Immediately, this implies that $\beta^{3/4} \leq n^{1/10}$, so we have
        \[
            n^{4/15}\beta^{5/4} \leq n^{7/15}\beta^{1/2}.
        \]
        Likewise, from \cref{rmk:gammabeta}, we know that $\gamma = O(\sqrt{\beta})$, so this means
        \[
         \frac{n}{\gamma} = \Omega \left (\frac{n}{\beta^{1/2}} \right)= \Omega( n^{14/15}) \geq  n^{8/15}.
        \]
        Again, we have two cases. First, if $\tau \leq n^{7/15}\beta^{1/2}$, then 
        \[
        \frac{\tau}{\beta^{1/2}} \leq n^{7/15}.
        \]
        Then, we can see 
        \[
        \frac{\mathbf{Progress}_1(\beta, \tau, \gamma)}{\mathbf{Rounds}_1(\beta, \tau, \gamma)} = \widetilde\Omega\left(\frac{n\beta}{\tau\gamma}\right) = \widetilde \Omega \left ( \frac{n \beta^{1/2}}{\tau} \right) = \widetilde \Omega \left ( n^{8/15}\right),
        \]
        where in the second equality we have used that $\gamma = O(\sqrt{\beta})$ as per \cref{rmk:gammabeta}. Otherwise, if $\tau \geq n^{4/15}\beta^{5/4}$, this means that 
        \[
        \frac{\tau^2}{\beta^{5/2}} \geq n^{8/15}.
        \]
        In this case, we see that 
        \[
        \frac{\mathbf{Progress}_4(\beta, \tau, \gamma)}{\mathbf{Rounds}_4(\beta, \tau, \gamma)} = \widetilde \Omega \left ( \min\left\{\frac{n}{\gamma},\frac{\tau^2}{\beta^2}\right\} \right) = \widetilde\Omega(n^{8/15}).
        \]
\end{enumerate}
Thus, in every case, we see that there is some choice of sub-rountine which guarantees a progress to round ratio of $\widetilde\Omega(n^{8/15})$.
\end{proof}

Finally, we can conclude with our main theorem:

\begin{theorem}
There is a randomized algorithm that, with high probability, finds a basis of any $n$-element matroid $\M$ in $\tilde{O}(n^{7/15})$ adaptive rounds, using only polynomially many independence queries per round.
\end{theorem}

\begin{proof}
    We simply run \cref{alg:earlyStop}, recovering the parameters $\gamma, \beta, \tau$ and sets $S_1, \dots S_{k}$. If the algorithm does not return on \cref{line:alpha-return}, we then invoke \cref{lem:caseBreakdown}: based on the values of $\beta, \gamma, \tau$ we can choose which subroutine to run. The result is that we have invested some $\kappa\geq 1$ adaptive rounds (and using $\mathrm{poly}(n)$ queries in each round), but by \cref{lem:caseBreakdown}, we are guaranteed to have made at least $\widetilde{\Omega}(\kappa \cdot n^{8/15})$ progress. In particular, this means that we have either recovered $\widetilde{\Omega}(\kappa \cdot n^{8/15})$ redundant elements (which we simply delete), or we have recovered $\widetilde{\Omega}(\kappa \cdot n^{8/15})$ independent elements, which we then contract on. In either case, we reduce the problem of computing a basis of $\M$ on $n$ elements, to computing a basis on a matroid with $n - \widetilde{\Omega}(\kappa \cdot n^{8/15})$ elements.

    On the other hand, if the algorithm returns on \cref{line:alpha-return}, by \cref{clm:return-recover}, we can recover an independent set of size $\widetilde \Omega(n)$ and contract. As we have invested $\kappa = O(k)=O(n^{1/3})$ rounds by \cref{clm:peel-round}, we also reduce the problem of computing a basis of $\M$ on $n$ elements, to computing a basis on a matroid with $n-\widetilde \Omega(n) \leq n-\widetilde \Omega(\kappa \cdot n^{8/15})$ elements.

    Thus, the total number of adaptive rounds required (denoted $T(n)$) obeys the recurrence
    \[
    T(n) \leq \kappa + T\left(n - \widetilde{\Omega}(\kappa \cdot n^{8/15})\right).
    \]
    In particular, this means that $T(n) = \widetilde{O}(n^{7/15})$ adaptive rounds suffice, as we desire. Note that the failure probability in each subroutine is exponentially small $1 / 2^{\Omega(n)}$ (and can even be amplified by repeating each subroutine $\mathrm{poly}(n)$ times in parallel), and so there is no concern of cascading errors becoming too large through rounds. Likewise, via \cref{clm:boundedErrorEstimation}, the error probability of our decomposition is also exponentially small. This yields the theorem. 
\end{proof}

\section{Conclusions}

We have presented the first progress in nearly four decades on the parallel complexity of finding bases in general matroids under independence-oracle access. Our main result improves the longstanding $O(\sqrt{n})$ upper bound of~\cite{kuw85}, achieving an $\tilde{O}(n^{7/15})$-round algorithm with polynomial query complexity.
As a corollary, we obtain a new upper bound for matroid intersection by integrating our techniques into the reduction of~\cite{BT25}. We also match the $\Omega(n^{1/3})$ lower bound for partition matroids by giving an $\tilde{O}(n^{1/3})$-round algorithm for this class.

Our results are achieved through a new matroid decomposition framework, a probabilistic analysis of rank deficiency via random sampling, and efficient parallel routines for contraction and deletion. We believe these techniques will pave the way for further algorithmic developments in the independence-oracle model for matroids.

\bibliographystyle{alpha}
\bibliography{ref}

@article{doe20,
  title={Probabilistic tools for the analysis of randomized optimization heuristics},
  author={Doerr, Benjamin},
  journal={Theory of evolutionary computation: Recent developments in discrete optimization},
  pages={1--87},
  year={2020},
  publisher={Springer}
}

@article{ehm91,
  title={Binomial approximation to the Poisson binomial distribution},
  author={Ehm, Werner},
  journal={Statistics \& Probability Letters},
  volume={11},
  number={1},
  pages={7--16},
  year={1991},
  publisher={Elsevier}
}

@inproceedings{Lov79,
  author       = {L{\'{a}}szl{\'{o}} Lov{\'{a}}sz},
  editor       = {Lothar Budach},
  title        = {On determinants, matchings, and random algorithms},
  booktitle    = {Fundamentals of Computation Theory, {FCT} 1979, Proceedings of the
                  Conference on Algebraic, Arthmetic, and Categorial Methods in Computation
                  Theory, Berlin/Wendisch-Rietz, Germany, September 17-21, 1979},
  pages        = {565--574},
  publisher    = {Akademie-Verlag, Berlin},
  year         = {1979},
  bibsource    = {dblp computer science bibliography, https://dblp.org}
}

@article{KUW86,
  author       = {Richard M. Karp and
                  Eli Upfal and
                  Avi Wigderson},
  title        = {Constructing a perfect matching is in random {NC}},
  journal      = {Comb.},
  volume       = {6},
  number       = {1},
  pages        = {35--48},
  year         = {1986},
  url          = {https://doi.org/10.1007/BF02579407},
  doi          = {10.1007/BF02579407},
  bibsource    = {dblp computer science bibliography, https://dblp.org}
}

@article{KUW88,
  author       = {Richard M. Karp and
                  Eli Upfal and
                  Avi Wigderson},
  title        = {The Complexity of Parallel Search},
  journal      = {J. Comput. Syst. Sci.},
  volume       = {36},
  number       = {2},
  pages        = {225--253},
  year         = {1988},
  url          = {https://doi.org/10.1016/0022-0000(88)90027-X},
  doi          = {10.1016/0022-0000(88)90027-X},
  bibsource    = {dblp computer science bibliography, https://dblp.org}
}

@inproceedings{ST17,
  author       = {Ola Svensson and
                  Jakub Tarnawski},
  editor       = {Chris Umans},
  title        = {The Matching Problem in General Graphs Is in Quasi-NC},
  booktitle    = {58th {IEEE} Annual Symposium on Foundations of Computer Science, {FOCS}
                  2017, Berkeley, CA, USA, October 15-17, 2017},
  pages        = {696--707},
  publisher    = {{IEEE} Computer Society},
  year         = {2017},
  url          = {https://doi.org/10.1109/FOCS.2017.70},
  doi          = {10.1109/FOCS.2017.70},
  bibsource    = {dblp computer science bibliography, https://dblp.org}
}

@article{Lub86,
  author       = {Michael Luby},
  title        = {A Simple Parallel Algorithm for the Maximal Independent Set Problem},
  journal      = {{SIAM} J. Comput.},
  volume       = {15},
  number       = {4},
  pages        = {1036--1053},
  year         = {1986},
  url          = {https://doi.org/10.1137/0215074},
  doi          = {10.1137/0215074},
  bibsource    = {dblp computer science bibliography, https://dblp.org}
}

@inproceedings{BS20,
  author       = {Eric Balkanski and
                  Yaron Singer},
  editor       = {Konstantin Makarychev and
                  Yury Makarychev and
                  Madhur Tulsiani and
                  Gautam Kamath and
                  Julia Chuzhoy},
  title        = {A lower bound for parallel submodular minimization},
  booktitle    = {Proceedings of the 52nd Annual {ACM} {SIGACT} Symposium on Theory
                  of Computing, {STOC} 2020, Chicago, IL, USA, June 22-26, 2020},
  pages        = {130--139},
  publisher    = {{ACM}},
  year         = {2020},
  url          = {https://doi.org/10.1145/3357713.3384287},
  doi          = {10.1145/3357713.3384287},
  bibsource    = {dblp computer science bibliography, https://dblp.org}
}

@inproceedings{CCK21,
  author       = {Deeparnab Chakrabarty and
                  Yu Chen and
                  Sanjeev Khanna},
  title        = {A Polynomial Lower Bound on the Number of Rounds for Parallel Submodular
                  Function Minimization},
  booktitle    = {62nd {IEEE} Annual Symposium on Foundations of Computer Science, {FOCS}
                  2021, Denver, CO, USA, February 7-10, 2022},
  pages        = {37--48},
  publisher    = {{IEEE}},
  year         = {2021},
  url          = {https://doi.org/10.1109/FOCS52979.2021.00013},
  doi          = {10.1109/FOCS52979.2021.00013},
  bibsource    = {dblp computer science bibliography, https://dblp.org}
}

@inproceedings{GGR22,
  author       = {Sumanta Ghosh and
                  Rohit Gurjar and
                  Roshan Raj},
  editor       = {Joseph (Seffi) Naor and
                  Niv Buchbinder},
  title        = {A Deterministic Parallel Reduction from Weighted Matroid Intersection
                  Search to Decision},
  booktitle    = {Proceedings of the 2022 {ACM-SIAM} Symposium on Discrete Algorithms,
                  {SODA} 2022, Virtual Conference / Alexandria, VA, USA, January 9 -
                  12, 2022},
  pages        = {1013--1035},
  publisher    = {{SIAM}},
  year         = {2022},
  url          = {https://doi.org/10.1137/1.9781611977073.44},
  doi          = {10.1137/1.9781611977073.44},
  bibsource    = {dblp computer science bibliography, https://dblp.org}
}

@article{BPV15,
  title={On the number of matroids},
  author={Bansal, Nikhil and Pendavingh, Rudi A and Van der Pol, Jorn G},
  journal={Combinatorica},
  volume={35},
  pages={253--277},
  year={2015},
  publisher={Springer}
}

@inproceedings{CGJS22,
  author       = {Deeparnab Chakrabarty and
                  Andrei Graur and
                  Haotian Jiang and
                  Aaron Sidford},
  title        = {Improved Lower Bounds for Submodular Function Minimization},
  booktitle    = {63rd {IEEE} Annual Symposium on Foundations of Computer Science, {FOCS}
                  2022, Denver, CO, USA, October 31 - November 3, 2022},
  pages        = {245--254},
  publisher    = {{IEEE}},
  year         = {2022},
  url          = {https://doi.org/10.1109/FOCS54457.2022.00030},
  doi          = {10.1109/FOCS54457.2022.00030},
  bibsource    = {dblp computer science bibliography, https://dblp.org}
}

@inbook{Qua23,
author = {Kent Quanrud},
title = {Quotient sparsification for submodular functions},
booktitle = {Proceedings of the 2024 Annual ACM-SIAM Symposium on Discrete Algorithms (SODA)},
chapter = {},
pages = {5209-5248},
	year         = {2024},
publisher    = {{SIAM}},
doi = {10.1137/1.9781611977912.187},
URL = {https://epubs.siam.org/doi/abs/10.1137/1.9781611977912.187},
eprint = {https://epubs.siam.org/doi/pdf/10.1137/1.9781611977912.187}
}

@inproceedings{Kar93,
  author       = {David R. Karger},
  editor       = {Vijaya Ramachandran},
  title        = {Global Min-cuts in \textbf{RNC}, and Other Ramifications of a Simple Min-Cut
                  Algorithm},
  booktitle    = {Proceedings of the Fourth Annual {ACM/SIGACT-SIAM} Symposium on Discrete
                  Algorithms, 25-27 January 1993, Austin, Texas, {USA}},
  pages        = {21--30},
  publisher    = {{ACM/SIAM}},
  year         = {1993},
  url          = {http://dl.acm.org/citation.cfm?id=313559.313605},
  bibsource    = {dblp computer science bibliography, https://dblp.org}
}

@inproceedings{FGT16,
  author       = {Stephen A. Fenner and
                  Rohit Gurjar and
                  Thomas Thierauf},
  editor       = {Daniel Wichs and
                  Yishay Mansour},
  title        = {Bipartite perfect matching is in quasi-NC},
  booktitle    = {Proceedings of the 48th Annual {ACM} {SIGACT} Symposium on Theory
                  of Computing, {STOC} 2016, Cambridge, MA, USA, June 18-21, 2016},
  pages        = {754--763},
  publisher    = {{ACM}},
  year         = {2016},
  url          = {https://doi.org/10.1145/2897518.2897564},
  doi          = {10.1145/2897518.2897564},
  bibsource    = {dblp computer science bibliography, https://dblp.org}
}

@inproceedings{KUW85,
  author       = {Richard M. Karp and
                  Eli Upfal and
                  Avi Wigderson},
  title        = {The Complexity of Parallel Computation on Matroids},
  booktitle    = {26th Annual Symposium on Foundations of Computer Science, Portland,
                  Oregon, USA, 21-23 October 1985},
  pages        = {541--550},
  publisher    = {{IEEE} Computer Society},
  year         = {1985},
  url          = {https://doi.org/10.1109/SFCS.1985.57},
  doi          = {10.1109/SFCS.1985.57},
  bibsource    = {dblp computer science bibliography, https://dblp.org}
}

@book{Oxl06,
  title={Matroid theory},
  author={Oxley, James G},
  volume={3},
  year={2006},
  publisher={Oxford University Press, USA}
}

@inproceedings{BT25,
  author       = {Joakim Blikstad and
                  Ta{-}Wei Tu},
  editor       = {Ioana Oriana Bercea and
                  Rasmus Pagh},
  title        = {Efficient Matroid Intersection via a Batch-Update Auction Algorithm},
  booktitle    = {2025 Symposium on Simplicity in Algorithms, {SOSA} 2025, New Orleans,
                  LA, USA, January 13-15, 2025},
  pages        = {226--237},
  publisher    = {{SIAM}},
  year         = {2025},
  url          = {https://doi.org/10.1137/1.9781611978315.18},
  doi          = {10.1137/1.9781611978315.18},
  bibsource    = {dblp computer science bibliography, https://dblp.org}
}

@inproceedings{Bli22,
  author       = {Joakim Blikstad},
  editor       = {Mikolaj Bojanczyk and
                  Emanuela Merelli and
                  David P. Woodruff},
  title        = {Sublinear-Round Parallel Matroid Intersection},
  booktitle    = {49th International Colloquium on Automata, Languages, and Programming,
                  {ICALP} 2022, July 4-8, 2022, Paris, France},
  series       = {LIPIcs},
  volume       = {229},
  pages        = {25:1--25:17},
  publisher    = {Schloss Dagstuhl - Leibniz-Zentrum f{\"{u}}r Informatik},
  year         = {2022},
  url          = {https://doi.org/10.4230/LIPIcs.ICALP.2022.25},
  doi          = {10.4230/LIPICS.ICALP.2022.25},
  bibsource    = {dblp computer science bibliography, https://dblp.org}
}

@inproceedings{GT17,
  title={Linear matroid intersection is in quasi-NC},
  author={Gurjar, Rohit and Thierauf, Thomas},
  booktitle={Proceedings of the 49th Annual ACM SIGACT Symposium on Theory of Computing},
  pages={821--830},
  year={2017}
}

\appendix

\section{Proof of \cref{thm:matroidIntersectionIntro}}\label{sec:matroidIntersection}

To start, we recall the following lemma of \cite{BT25}:

\begin{lemma}[\cite{BT25}, Fact 2.4 and Corollary 3.8]
Let $\M_1=(E,\I_1),\M_2=(E,\I_2)$ be 2 matroids. Let $n=|E|$ and $r$ be the size of the size of the largest independent set of $\M_1,\M_2$. Then
\begin{itemize}
    \item There is an $\widetilde O\left(\frac{nT(n)}{\varepsilon\Delta}\right)$ rounds independence-query algorithm that finds a common independent set $S\in \I_1\cap \I_2$ of size $|S|\geq r-(\varepsilon r+\Delta)$, given that there is a $T(n)$ round independence-query algorithm that finds a maximum weight basis of a matroid on $n$ elements.
    \item Given $S\in\I_1 \cap \I_2$, in a single round of independence query, one can compute an $S'\in \I_1 \cap \I_2$ of size $|S'|=|S|+1$ or decide that $S$ is of maximum possible size.
\end{itemize}
\end{lemma}

Now, we show \cref{thm:matroidIntersectionIntro}:

\begin{theorem}[see also \cite{BT25} Theorem 1.4][\cref{thm:matroidIntersectionIntro} restated]
There is a randomized algorithm that, with high probability, finds a maximum common independent set of two $n$-element matroids in $\tilde{O}(n^{37/45})$ adaptive rounds, using only polynomially many independence queries per round.
\end{theorem}
\begin{proof}
First, as in Lemma 2.2 of \cite{BT25}, observe that an $r$-round independence query algorithm for computing the basis of an $n$ element matroid immediately yields an $r$ round algorithm for computing a \emph{maximum weight} basis of an $n$ element matroid. Thus, we can use our $\widetilde{O}(n^{7/15})$ round algorithm to also find maximum weight bases.

To proceed, we thenWe set $\varepsilon = n^{22/45}r^{-2/3}$ and $\Delta = \varepsilon r= n^{22/45}r^{1/3}$. We can first find an $S\in \I_1\cap \I_2$ of size $|S|\geq r-(\varepsilon r +\Delta)$ in
\[
\widetilde O\left( \frac{n\cdot n^{7/15}}{\varepsilon \Delta} \right) = \widetilde O(n^{22/45} r^{1/3})
\]
rounds and then augment it to optimal in $O(\varepsilon r+\Delta)=O(n^{22/45}r^{1/3})$ rounds. As $r\leq n$, the total rounds of adaptivity needed is $\widetilde O(n^{37/45})$.
\end{proof}

\end{document}